\documentstyle[10pt, aaspp4]{article}

\def\mol#1#2#3{$^{#1}{\rm #2(#3)}$}

\begin{document}

\title{Simultaneous Observations of GRS 1758--258 in 1997 
by VLA, IRAM, SEST, RXTE and OSSE: Spectroscopy and Timing }
\author{D. Lin\altaffilmark{1}, I. A. Smith\altaffilmark{1}, 
E. P. Liang\altaffilmark{1}, T. Bridgman\altaffilmark{2}, 
D. M. Smith\altaffilmark{3}, J. Mart\'{\i}\altaffilmark{4}, 
Ph. Durouchoux\altaffilmark{5}, I.F. Mirabel\altaffilmark{5, 6}, \&
L.F. Rodr\'{\i}guez\altaffilmark{7}}

\altaffiltext{1}{Department of Space Physics and Astronomy, Rice University, 
6100 S. Main, Houston, TX 77005, USA. Email: lin@spacsun.rice.edu}

\altaffiltext{2}{Goddard Space Flight Center, Code 660.1, Greenbelt, MD 20771}

\altaffiltext{3}{Space Sciences Laboratory, University of California Berkeley, 
Berkeley, CA 94720}

\altaffiltext{4}{Universidad de Ja\'en, Departamento de F\'{\i}sica,
Escuela Polit\'ecnica Superior, Calle Virgen de la Cabeza, 2,
E-23071 Ja\'en, Spain}

\altaffiltext{5}{CEA/DSM/DAPNIA Service d'Astrophysique, F-91191,
Gif-sur-Yvette, France}

\altaffiltext{6}{Instituto de Astronom\'{\i}a y F\'{\i}sica del Espacio,
Casilla de Correo 67, Sucursal 28, Buenos Aires 1428, Argentina.}

\altaffiltext{7}{Instituto de Astronom\'{\i}a, UNAM, J.J. Tablada 1006,
Morelia, Michoac\'an, 58090, M\'exico}

\begin{abstract}
We report the results of our multi-wavelength observations of GRS 1758--258 made 
in August 1997. The energy bands include radio, millimeter, X-ray, and gamma-ray.
The observations enable us to obtain a complete spectrum of the source over an 
energy range of 2 -- 500 keV. The spectrum shows that GRS 1758--258
was in its hard state.  It is well fitted by the Sunyaev-Titarchuk (ST) Compton
scattering model with a plasma temperature of 45 keV and a Thomson depth of 3.3.
Taking relativistic effects into account, we get a little higher plasma temperature 
(52 keV) by using the improved version of the ST model (HT model) plus a soft
black-body component. The spectrum is also fit by a power law with an 
exponential cutoff (PLE) plus a soft black-body component. The temperature 
of the soft components in both models is about 1.2 keV, and the energy flux 
is less than 1.5\% of the total X- and gamma-ray flux. The deduced hydrogen 
column density is in the range of $(0.93 - 2.0) \times 10^{22}$ cm$^{-2}$.
No significant iron lines are detected. The radio emission has a flat energy
spectrum. The daily radio, X-ray and gamma-ray light curves show that 
GRS 1758--258 was stable during the observation period, but was highly 
variable on smaller time scales in X- and gamma-rays. The power density 
spectra are typical for the low-state, but we find the photon flux for 
the 5 to 10 keV band to be more variable than that in the other two energy 
bands (2 -- 5 keV and 10 -- 40 keV). Harmonically spaced quasi-periodic 
oscillations (QPOs) are observed in the power spectra. The phase lags 
between the hard photons and the soft photons have a flat distribution 
over a wide range of frequencies. A high coherence of about 1.0 
(0.01 -- 1 Hz) between the hard photons and the soft photons is also 
obtained in our observations. We compare these results with two variation 
models. Our millimeter observations did not reveal any conclusive signatures 
of an interaction between the jet from GRS 1758--258 and the molecular 
cloud that lies  in the direction of GRS 1758--258.
\end{abstract}


\section{Introduction}

GRS1758--258, discovered by ART-P and SIGMA on GRANAT in the spring 
of 1990, is one of the brightest persistent sources above 100 keV  in the 
vicinity of the galactic center (\cite{sunyaev91}). Its spectrum extends up 
to 300 keV, and because this hard spectrum resembles those of other black 
hole candidates (BHC) in the hard (= soft X-ray low) state, GRS1758--258 
is generally considered to be a BHC (\cite{tanaka95}; \cite{ln97};  
\cite{liang98}; \cite{poutanen98}). The nature of the donor star orbiting 
around the black hole is still controversial, although a high mass 
companion is certainly ruled out (e.g. \cite{marti98}). Monitoring has 
shown that the source was  in this hard state for almost a year during 
a three-year period before the hard X-ray flux (above 35 keV) decreased 
by a factor of 10  (\cite{gilfanov93}; \cite{kuznetsou99}). This decrease 
in the hard X-ray flux may indicate a transition from the hard state to 
the soft (= soft X-ray high) state. Unfortunately, the lack of 
simultaneous soft X-ray observations prevents us from drawing a definitive 
conclusion about the transition. A significant soft X-ray excess
was suggested by simultaneous ROSAT-SIGMA observations in 1993, when
the hard component had a lower intensity (\cite{mereghetti94}). But 
a reanalysis suggests that the excess was due to the contamination 
from GX 5--1 (\cite{gps97}). From ASCA observations, it was inferred 
that the soft component contributes only a few percent of the total 
X-ray and gamma-ray flux when the source is in its hard state although 
there were no simultaneous gamma-ray observations (\cite{mereghetti97}). 
A significant soft component for GRS 1758--258 was also reported by 
\cite{hs98}, and in 1999 it has oscillated between the hard and soft 
states(\cite{shs99}). To be able to understand the nature of the 
soft component when the spectrum is in its hard state, it is essential to have 
good simultaneous coverage of GRS 1758--258 over as broad an energy range 
as possible. 

Here we report the results of our simultaneous radio, millimeter, X-ray 
and gamma-ray observations of GRS 1758--258 in August 1997. The radio 
observations were performed by the Very Large Array (VLA). The millimeter 
observations were performed by IRAM, and we compared to the early 
observations made using the Swedish ESO Sub millimeter Telescope (SEST)
at La Silla (Chile). The X-ray and gamma-ray observations 
involved the Proportional Counter Array (PCA) and High Energy X-ray 
Timing Experiments (HEXTE) on the {\it Rossi X-Ray Timing Explorer 
(RXTE)} and the Oriented Scintillation Spectrometer Experiment (OSSE) 
on the {\it Compton Gamma-Ray Observatory (CGRO)}. Since the energy 
ranges of these three instruments overlap, we get a good measurement 
of the complete X- and gamma-ray spectrum.  

The rapid X-ray variability can also provide direct information about 
the accretion disk. Observations of GRS 1758--258 made by {\it RXTE} 
in 1996 (\cite{dsmith97}) found that the source had a typical power 
spectrum for BHCs in their hard states. The integrated fractional rms 
amplitude (IFRA) was 24\%. Unlike the case in 1996 when the power 
spectrum had little energy dependence, our observations show that 
the 5 -- 10 keV band was more variable than the 2 -- 5 keV and 
10 -- 40 keV bands.

In section 2, we describe the details of the observations and data 
reductions. Daily and short time scale light curves are shown in 
section 3. In section 4, we present the analysis of the spectral 
data. The timing analysis is discussed in section 5. We summarize 
the results in section 6.

\section{Observations and Data Reductions}
 
\subsection{Radio Observations}

At radio wavelengths, GRS 1758--258 has been found to be a ``microquasar''
with relativistic jets emanating from a central variable source VLA-C
(\cite{rmm92}; \cite{mr94}).

We performed a series of nine VLA observations between 1997 August 3 -- 24,
spanning our entire {\it CGRO} run. Some of these were simultaneous 
with the {\it RXTE} observations discussed in \S 2.2. All the 
observations were made in the C configuration at the 6 and 3.5 cm 
wavelengths (4860 and 8460 MHz), with a bandwidth of 100 MHz.
The VLA runs were all 2 hours long except for 1.5 hours on August 15.
This included time for both amplitude and phase calibration: the 
amplitude calibrator was 1331+305 and the phase calibrator 1751--253, 
whose bootstrapped flux density was always close to 0.48 Jy at 6 cm 
and 0.27 Jy at 3.5 cm, respectively.

Table 1 gives the flux densities for the central radio source for each of
our separate observations. The flux densities and errors have been 
measured by fitting an elliptical Gaussian to the source with the 
AIPS task JMFIT. The upper limits are $4 \sigma$ based on the map 
rms noise, measured using the AIPS task IMSTA. The radio light curves 
are also plotted in Figure 1. As has been found previously, VLA-C is 
often a weak radio source close to the detection threshold in our 
daily observations.

The central source is clearly detected ($>10 \sigma$) when we 
concatenate the VLA data from all our 1997 August 3--24 observations.
The average flux density values were $0.14 \pm 0.01$ mJy at 4860 MHz 
and $0.14 \pm 0.01$ mJy at 8460 MHz, giving a flat spectral index of 
$0.0 \pm 0.1$. We combine the averaged radio spectrum with our 
averaged X-ray and gamma-ray spectra in Figure 8.

The radio emission suggests the continual presence of an outflow 
from VLA-C. The approximately flat spectrum is not consistent with 
optically thin emission (unless the electron distribution is 
exceptionally hard) and is indicative of some synchrotron 
self-absorption in the radio-emitting regions. One possible geometry 
would be a conical jet such as discussed in
\cite{hj88}.

For comparison, in 1997 August the nearby source VLA-D had average flux
densities of $0.14 \pm 0.01$ mJy at 4860 MHz and $0.07 \pm 0.01$ mJy at
8460 MHz, giving a spectral index of $-1.3 \pm 0.2$.
The 6 cm flux density of VLA-D is $\sim 1/2$ the value it had in 1992
which was also measured in the C configuration (\cite{rmm92}).
Maps and further details of these and other VLA observations of GRS 1758--258
will be presented elsewhere (J. Mart\'{\i} et al. 2000, in preparation).

\subsection{Millimeter Observations}

The discovery of a molecular cloud 
(Bally \& Leventhal 1991; Mirabel et al. 1991)
in the direction of 1E 1740.7--2942 raised the picture of a compact source
embedded inside a molecular cloud that is accreting material from the cloud.
GRS 1758--258 has similar X-ray properties to 1E 1740.7--2942, but the
lower value of $n_{\rm H}$ $\sim 2 \times 10^{22}~{\rm cm}^{-2}$ 
suggests that GRS 1758--258 is not inside a dense
molecular cloud (Chen, Gehrels, \& Leventhal 1994).

To determine whether there is a molecular cloud in the direction of
GRS 1758--258, we performed millimeter observations during our 
multi-wavelength campaign on 1997 August 7 using the IRAM 30-m telescope 
at Pico Veleta (Spain). We also report here on earlier millimeter 
observations of GRS 1758--258 that were conducted in April 1995 and 
October 1995 using the 15-m SEST.

As well as observing the core radio source, we also made millimeter
observations along the radio jets in order to look for signatures of 
the jet interaction with a molecular cloud such as an increase in the 
width of the line or the detection of wings in the peak. For example, 
we previously detected a correlation between the ROSAT X-ray
hot spots and the CO mm measurements in observations of SS 433 
(Durouchoux et al. 1999). The higher ionization environments in 
shocked molecular gas and near X-ray sources can also enhance 
the emission from molecules such as ${\rm HCO}^+$ and HCN (Krolik 
\& Kallman 1983; Phillips \& Lazio 1995; Lepp \& Dalgarno 1996;
Yan \& Dalgarno 1997).

\subsubsection{SEST Observations}

We performed observations using the \mol{12}{CO}{1-0} transition at 115.3 GHz
and the \mol{12}{CO}{2-1} transition at 230.5 GHz. Observations 
were made in the direction of the core source ($\alpha =$ 17:58:07, 
$\delta = -$ 25:44:30, epoch 1950.0) and at one location in the 
northern radio jet ($\alpha =$ 17:58:06.8, $\delta = -$ 25:43:45)
and one in the southern jet ($\alpha =$ 17:58:05.5, $\delta = -$ 25:45:50).
The observations were carried out in a position beam switching mode, choosing 
an absolute reference position $-15\arcmin$ in $\alpha$ and $+20\arcmin$ in 
$\delta$ away from the source. Such a distant location was required 
to find a region without radio emission in the crowded Galactic center area.

The receivers yielded an overall system temperature (including sky noise)
$\sim 400$ and 300 K at 115 and 230 GHz respectively. The backend was 
an acousto-optical spectrometer. The low resolution spectrometer 
with a channel separation of 0.7 MHz (or 2.4 ${\rm km~s}^{-1}$) was 
used for these observations. The antenna beam-widths were 45\arcsec\ 
and 23\arcsec\ for the \mol{12}{CO}{1-0} and \mol{12}{CO}{2-1} 
transitions respectively. All the spectral intensities were converted 
to the main brightness temperature scale, correcting for a 
mean beam efficiency of 0.7 at 115 GHz and 0.5 at 230 GHz. 
The calibration was checked by monitoring M17SW, and was found to be 
consistent between the different observing periods to within 19\%. 
The pointing accuracy, obtained from measurements of the SiO 
maser W Hya, was better than 5\arcsec.

A molecular cloud was found in \mol{12}{CO}{1-0} and \mol{12}{CO}{2-1} in 
the direction of GRS 1758--258 with a velocity $\sim 75 ~{\rm km~s}^{-1}$.
The results are given in Table 2.
The peak area \mol{12}{CO}{2-1}/\mol{12}{CO}{1-0} ratio of $\sim 0.5$ is 
consistent with the values found in other molecular clouds (Oka et al. 1998).

A \mol{12}{CO}{1-0} line with the same velocity was also detected in the 
northern and southern parts of the jet.
For the northern direction, the flux density was slightly lower than in 
the core direction, but not significantly.
For the southern direction, the width of the line is slightly larger than 
for the core source, but again not significantly.
\mol{12}{CO}{2-1} was also detected in the north and south jets with
velocities, peak areas, and line widths consistent with the ones measured 
for this transition in the core source.
We conclude that there are no obvious signatures for the interaction
of the jet with the molecular cloud.

We also searched for the $75 ~{\rm km~s}^{-1}$ line in the 
\mol{13}{CO}{1-0} and \mol{13}{CO}{2-1} transitions, but no significant
detections were made: integrating over 10 minutes the mean $1 \sigma$ upper 
limit was $\sim 0.8$ K for both the transitions.
We also did not detect the \mol{}{HCN}{1-0} or \mol{}{HCN}{2-1} lines: 
the mean $1 \sigma$ upper limit was $\sim 0.05$ K.

\subsubsection{IRAM Observations}

We observed simultaneously in the \mol{}{HCO^+}{1-0}
transition at 89.2 GHz with the SIS 3mm-1 receiver, 
in the \mol{}{HCN}{1-0} transition at 88.6 GHz with the SIS 3mm-2 receiver, 
in the \mol{13}{CO}{2-1} at 220.4 GHz with the 230G1 receiver,
and the \mol{12}{CO}{2-1} transition at 230.5 GHz with the 230G2 receiver.
The half power beams varied from 27\arcsec\ at 90 GHz to 10.4\arcsec\ at 
230 GHz.
As for the SEST runs, the observations were carried out in a position beam 
switching mode, choosing an absolute reference position $-15\arcmin$ in 
$\alpha$ and $+20\arcmin$ in $\delta$ away from the source.
The main system temperatures were 356, 389, 912, and 1000 K, and the opacities 
0.16, 0.18, 0.58, and 0.62 at 89.2, 88.6, 220.4, and 230.5 GHz respectively. 

We observed GRS 1758--258 at 12 different positions on the jet axis
separated by 24\arcsec.
The integration time was 1 minute on each location.
The individual observations did not have sufficient significance to
study the line emissions, and so we added them together to make an
average result for each transition.

We did not detect any lines in the \mol{}{HCO^+}{1-0} (0.04 K, $1 \sigma$)
and \mol{13}{CO}{2-1} (0.6 K, $1 \sigma$) spectra.
The \mol{12}{CO}{2-1} transition was clearly detected at
$73.81 \pm 0.48 ~{\rm km~s}^{-1}$ with a line width of 
$11.19 \pm 1.29 ~{\rm km~s}^{-1}$, similar to that seen earlier by SEST.
However, the intensity of $1.32 \pm 0.2$ K and
peak area of $15.7 \pm 1.42 ~{\rm K~km~s}^{-1}$ are almost double
those seen earlier by SEST.

The \mol{}{HCN}{1-0} transition was also detected in our IRAM observation at 
$76.85 \pm 0.67 ~{\rm km~s}^{-1}$ with a line width of 
$5.29 \pm 1.29 ~{\rm km~s}^{-1}$, intensity $0.14 \pm 0.04$ K
and peak area of $0.79 \pm 0.19 ~{\rm K~km~s}^{-1}$.
Although the statistical significance is weak, this also seems to be
brighter than was seen in our SEST observations.

The comparison between the SEST and IRAM results suggests a brightening
in the outflow in the 2 years between the observations.
The fact that the IRAM beam size is less than half that of SEST implies
that if the emission is diffuse, there has actually been an increase of
$\sim 4$ in the line emissions.
However, we caution that a longer-term monitoring program is required to
prove that the line emissions are variable.

We conclude that while there is a molecular cloud in the direction of
GRS 1758--258, we have not yet found any conclusive signatures that there 
is an interaction between the jet and this molecular cloud.
We suggest that longer observations using the \mol{}{SiO}{2-1} transition 
would also be worthwhile to search for an interaction.

\subsection{ RXTE Observations}

The PCA consists of five Xe proportional counters with a total effective 
area of about 6500 $ {\rm cm^{2}}$ (\cite{jahoda96}). Its sensitive energy range is 
from 2 to 60 keV with an energy resolution of $<$ 18\% at 6 keV, and a time 
resolution of 1 ${\rm \mu}$s. The HEXTE consists of two independent clusters 
(clusters A and B) each containing four phoswich scintillation detectors 
(\cite{rothschild98}). It has a total effective area of 1600 cm${\rm ^{2}}$
and a sensitive energy range of 15 -- 250 keV with an energy resolution of 
15 \% at  60 keV and a time resolution of 8 ${\rm \mu}$s. Both the PCA and 
the HEXTE have a field of view of $\approx {\rm 1^{\circ}}$ (FWHM). 

Both the PCA and HEXTE made observations on 5 days during the period of 
August 13 - 24, 1997. During the  observations, the  on-source pointing 
direction (RA = 270.315$^{\circ}$, DEC = --26.245$^{\circ}$, J2000) was 
offset from GRS 1758--258 by half a degree to avoid the X-ray source GX 5--1, 
which is $1.2^{\circ}$ away from the on-source location. 
The flux of GX 5--1 during the observation was 
about 1 Crab based on the observations by the All Sky Monitor (ASM) on 
the {\it RXTE}. However, the offsetting effectively reduced the contamination 
from GX 5--1, whose contribution is estimated to be much less than 0.033\% 
of the total count rate in the energy band of 2 -- 12 keV.  The HEXTE 
cluster A was rocked by $\pm 1.5^{\circ}$ and cluster B by 
$\pm 3.0^{\circ}$ to make the background measurements with a 
switching time of 16 seconds. The PCA has  no direct background 
measurements.  

The PCA and HEXTE data were reduced using FTOOLS 4.1.1.  Both data sets 
were screened so that only those data were used when the spacecraft was 
not passing through the South Atlantic Anomaly and when the source was 
observed at elevations ${\rm > 10^{\circ}}$ above the Earth's limb and the 
pointing was offset ${\rm <0.02^{\circ}}$ from the on-source pointing
to filter out the slew data. After the screening, 14 observation segments 
have valid data. The start and stop times of each segment are listed in 
Table 3. Figure 2 shows the PCA 16-second resolution light curves for the 14 
observation segments.

For the PCA, Standard 2 production data from all layers of the five PCU detectors
were used to generate an average spectrum over our entire {\it RXTE} run. 
The FTOOLS program PCARSP (version 2.36) was used to 
generate the response matrix file with 
corrections made for the off-axis pointings. The total exposure time with 
dead-time correction is 33399 seconds. A scheme of adding systematic 
errors to the PCA data was also applied (\cite{rothschild98}).  
The added errors are 0.5\% for 2.5 -- 5.5 keV, 1\% for 5.5 -- 8 keV, 0.5\% for
8 -- 20 keV, and 3\% for 20 -- 90 keV. The estimation of the PCA background
is complicated by the fact that the current PCA background models do not
include the galactic diffuse emission, which significantly contaminates our
observations. We used the previous {\it RXTE} observations in the 
vicinity of GRS 1758--258 (\cite{dsmith97}) to estimate the contribution 
from the diffuse emission. 
By combining these background data with the detector internal background 
(Q6 + activation), we are able to estimate the true background for our PCA
measurements. In the timing analysis, we neglect the time variation of the 
diffuse emission, which is estimated to be less than 10\% of that of GRS 1758--258
based on the background observations. Therefore, we remove the effects of
the diffuse emission by subtracting from the light curves constant count rates 
of 15.1 counts/s for the  2 -- 5 keV band, 17.9 counts/s for 5 -- 10 keV, and 
8.4 counts/s for 10 -- 40 keV.

For the HEXTE, we only used the cluster A data for the spectral analysis, because the 
data from cluster B are less reliable due to the count piling problem and 
the off-axis pointing. The data were taken using the E\_8us\_256\_DXIF mode.
The HEXTE background measurements agree very well with 
the on source measurement in the energy channels above 125 keV. The  response 
matrix file  hexte\_97mar20c\_pwa.rmf was used. The collimator effective area
has been corrected with the calibration file hexte\_pwa.fov for the off-axis pointing. 
The total exposure time with dead-time correction is 9683 seconds.

\subsection{OSSE Observations}

The OSSE instrument, consisting of four nearly identical NaI-CsI 
phoswich detectors, has a field of view  of 3.8$^{\circ}$ $\times$ 11.4$^{\circ}$ 
(FWHM), and has a sensitive energy range of 0.05 -- 10 MeV with an energy resolution 
of 7.8\% at 661 keV (\cite{johnson93}). 

The current OSSE observations were carried out on August 5 -- 19, 1997. 
During the  observations,  the four detectors made 
alternating on-source and off-source pointings with 2.048 minute dwell time 
on each pointing. The rocking direction, which is 
perpendicular to the long dimension of the collimator, was at 
an angle of 41$^{\circ}$ to the galactic plane to avoid confusing sources. 
The on-source pointing was at RA = 270.45$^{\circ}$, DEC = 26.12$^{\circ}$.
Detectors 1 and 2 did the standard ${\rm \pm12^{\circ}}$ off-source pointings 
(RA = 257.45$^{\circ}$, DEC = --29.59$^{\circ}$; RA = 282.5$^{\circ}$, 
DEC = --21.37$^{\circ}$). However, 
due to the instruments' configuration, detectors 3 and 4 could not be 
rocked to the regular off-source pointings, and  instead had off-source 
pointings at + 12$^{\circ}$  and  --7$^{\circ}$  (RA = 277.60$^{\circ}$, 
DEC = --23.48$^{\circ}$). Unfortunately, the known OSSE source GS 1826--24
(\cite{strickman96}) sits in the center of the --7$^\circ$ 
background pointing, and so we did not use the data from detectors 3 and 4.

The OSSE data were reduced with the IGORE data analysis package. 
A count spectrum and a response matrix were generated 
from the data of detectors 1 and 2 with the background subtracted.
Both the  count spectrum and the response matrix could be read by  
XSPEC (version 10.0) to make a joint fit with the PCA and the HEXTE data. 
The total OSSE exposure time is 320078 seconds.

Since the background pointings are 12 degrees away from the on-source pointing, 
the measured background, which IGORE automatically subtracts from the data 
during the processing, may  be different from the real background at 
the on-source location. We have checked all the sources monitored by 
the ASM/{\it RXTE}. No other significant hard X-ray point sources are in the fields of view
of detectors 1 and 2. However, an important problem is the spatial variation 
in the galactic diffuse emission (GDE). Since our background pointings are 
farther away from the galactic center than the on-source pointing, we need to 
make a correction to get the true on-source background. Based on several 
OSSE observations of the GDE and coordinated SIGMA observations 
(\cite{purcell96}), a model of the GDE distribution in and around the 
galactic plane is:
\begin{equation}
P(E, l, b) = [N  (\frac{E}{100 keV})^{-\alpha} + f(l)  {\rm posm}(E)] 
e^{-\frac{b^{2}}{b_{0}^{2}}}
\end{equation}
where $P$ is the photon flux emitted from a unit area in  the galactic coordinates, 
$E$ is the photon energy in  keV, $l$ and $b$ are the galactic longitude and
latitude respectively, posm($E$) is the positronium continuum spectrum 
defined in XPSEC, $f(l)$ is the normalization factor of posm($E$), $N$ = 0.78 
photons/cm$^{2}$/s/keV/degree$^{2}$, $\alpha$ = 2.55, $f(l=0^{\circ})$ = $1.1 
\times 10^{-2}$, $f(l=25^{\circ})$ = $f(l=339^{\circ})$ = $2.2 \times 10^{-3} $, 
and $b_{0}$ = $3.24^{\circ}$. These numbers were determined by folding 
the observations at l=0$^{\circ}$, 25$^{\circ}$ and 339$^{\circ}$  through 
the detector response function, which is assumed to be a Gaussian function 
with a two-dimension FWHM of 3.8$^{\circ}$ $\times$ 11.4$^{\circ}$. The lack 
of measurements at locations other than these three longitudes prevents us 
from determining the exact form of $f(l)$. But we can still set  upper and 
lower limits to the background correction that should be made to our OSSE 
data. The lower limit is when $f(l)$ = $f(l=25^{\circ})$. The upper limit is when  
$f(l)$ = $f(l=0^{\circ})$ for the on-source measurements and  $f(l)$ = 
$f(l=25^{\circ})$ for the off-source measurements. The two limiting versions 
of the correction spectrum are shown in Figure 4, which can be simulated 
with XSPEC and then subtracted from the OSSE data. 

\section{Light Curves}

Daily light curves of the VLA,  {\it RXTE}, and OSSE observations are shown 
in Figure 1. We fit constants to the light curves. The fitting results,
listed in Table 4, show that GRS 1758--258 was stable on the day time scale,
although there was a small decline in the flux observed by the PCA. 

On shorter time scales, GRS 1758--258 is clearly variable.
Significant variations are seen in the PCA 16-second light curves (Figure 2), which
show the 14 valid observation segments. Some small bursts are
seen in the light curves. In Figure 3, the observation segment 11 is
examined at higher time resolutions. The peak near 650 seconds 
in the 16-second resolution  light curve actually consists of several small 
bursts on the 1-second time scale.  Moreover, a narrow burst near 630 seconds 
in the 0.1-second light curve still has substructure 
in the 0.01-second resolution light curve. However, these bursts are 
relatively small compared to those in other sources such as
GX 339--4 (\cite{ismith99}).    

\section{Spectrum analysis}
For all three instruments, we combined all the data to make one spectrum. 
Though the source is highly variable on small time scales, we will show in 
section 5.3 that there are no significant trends in the hardness ratio. Thus 
the average spectrum shown here should be a good representative shape, 
though the instantaneous amplitude varies significantly with time.
Since the HEXTE data has fewer problems with the background, in the following 
discussion we will normalize  the OSSE  and  PCA spectra to the HEXTE spectrum.

The HEXTE data can be well fit by a power law with  photon index 
of 1.55$\pm$0.04  in the energy range of 15 -- 125 keV (Figure 5). Above 125 keV,
the background dominates the data. The photon flux between 20 -- 100 keV is 
(2.0 $\pm$ 0.4) $\times$ 10$^{-2}$ photons/cm$^{2}$s. This is consistent
with the average flux of  (2.8 $\pm$ 0.2)  $\times$ 10$^{-2}$  photons/cm$^{2}$s
detected by the BATSE Earth Occultation technique in the same period. 
The errors are 1-sigma throughout this paper.

Figure 6 shows the joint spectra of the OSSE and HEXTE data. The fit model
is a power law with an exponential cutoff (PLE):
\begin{equation}
F(E) = N_{0} E^{-\alpha}{\rm exp}(-\frac{E}{kT})
\end{equation}
where $ F(E)$ is the photon flux, $N_{0}$ is the normalization flux in the unit 
of photons/cm$^{2}$/s/keV, E is the photon energy in keV, $\alpha$ is 
the photon power-law index, and $kT$ is the cutoff energy. The higher curve in the 
figure, which has been scaled up by a factor of 10 for  illustrative purposes, 
has the minimum background correction applied to the OSSE data, while 
the lower curve is the maximum background correction case.  The most 
significant spectral difference between the minimum and maximum 
corrections is the difference in the cutoff energies. This is not surprising, 
since the effect of the positronium continuum is significant only when the energy 
is close to and less than 511 keV. 

However, for a joint fit to the PCA, HEXTE and OSSE data, the PLE model alone 
(Figure 7(e)) is not as good as the Sunyaev-Titarchuk (ST) Compton scattering 
model (\cite{st80}). An example of a ST fit is shown in Figure 7 (a) \& (b).
In Table  5, we give the fit parameters for two cases, one using the 
Maximum OSSE background correction, the other using the Minimum. Since these two extreme 
cases set the limits for the parameters, we use the average values over the two cases
(the Average column in Table 5) as the true values. Assuming a distance of 8.7 kpc, 
we get  a soft X-ray luminosity (2 -- 20 keV) of $1.17 \times 10^{37}$ ergs/s and a 
hard X-ray luminosity (20 -- 200 keV) of $3.09 \times 10^{37}$ ergs/s. These values
place GRS 1758--258 outside the ``X-ray burst box'' in Figure 1 of \cite{barret96}. 
This further supports that GRS 1758--258 is a black hole.  The hydrogen column density 
is (2.0 $\pm$ 0.1) $\times 10^{22}$ cm$^{-2}$, which is consistent with the ASCA and 
EXOSAT observations (\cite{shimer91}; \cite{mereghetti97}).  The residuals in Figure 7 
suggest that no significant iron lines are seen in the spectrum.

\cite{titarchuk94} and \cite{ht95} generalized the ST model to cover
a wide range of the plasma parameters, such as low Thomson depths
and relativistic effects. The best HT model fit to the data gives 
a negative  hydrogen column density ($n_{H} < 0$) and a seed photon 
energy $T_{0} = 0.85$ keV (or the cutoff energy in the lower end of 
the spectrum). The negative $n_{H}$ suggests that the true value of 
$T_{0}$ must be much smaller than 0.85 keV. In the further model fits,  
we set $T_{0}$ = 0.01 keV. because we don't have valid data below 2.5 keV, 
the exact value of $T_{0}$ does not affect the fit results as long as 
it is much smaller than 0.85 keV. With the fixed $T_{0}$, the HT model alone
can not fit the data well. Adding a back-body component to the model 
gives a good fit (Figure 7(c)). The best fit parameters are listed in
Table 6. Compared with the ST model, the HT model gives a slightly
higher plasma temperature due to the relativistic corrections. 
The black body temperature is about 1.18 keV, which is much higher
that the seed photon energy ($T_{0} < 0.85$ keV). Therefore, it is very
likely that the soft component and the seed photons have different
origins. The total flux of the soft black-body componnent is  
$3.46 \times 10^{-11}$ ergs/cm$^2$/s, which is about 0.9\% of 
the total continuum flux from 2 keV to 10 MeV. The low flux of 
the soft component in the hard state has been reported in the ASCA 
observations (\cite{mereghetti97}). The hydrogen column density given by 
the HT model is lower than that by the  ST model, but is close 
to the ASCA results (\cite{mereghetti97}).

Adding a soft component to the PLE model also improves the fit, in particular 
to the shape of the residuals (Figure 7(d)). The fit parameters are 
listed in Table 7. The soft black-body component (BB) has a temperature 
of $1.27\pm 0.06$ keV and a total flux of  $5.11 \times 10^{-11}$ 
ergs/cm$^2$/s, which is about 1.3 \% of the total continuum flux from 
2 keV to 10 MeV and is slightly higher than that given by the HT model. 
The hydrogen column density obtained from the (PLE + BB) model is lower 
than that from previous EXOSAT and  ASCA observations. We therefore 
conclude that the (PLE + BB) model is less favored than the Compton 
scattering models (ST and HT models).

Figure 7(f) shows that a power-law model cannot fit the data. This means
that the exponential cut-off is real. In Figure 8 we show the unfolded 
and normalized PCA, HEXTE and OSSE spectrum using the ST model.
 
\section{Timing Analysis}

The timing analysis was done only on the GoodXenon PCA data using all 
five PCUs. After the basic steps mentioned in section 2,  
background-subtracted light curves were generated for three energy 
bands: 2 -- 5 keV, 5 -- 10 keV, and 10 -- 40 keV. From these  light 
curves, power spectra, phase lags, coherences and hardness ratios
were calculated. In the calculations, the counts have been 
summed into 46.875 ms bins, the power spectra etc. were generated 
for intervals that are  384 seconds long, and the results were then 
averaged.

\subsection{Power Density Spectrum}
We performed the power density spectra (PDS) analysis on the 14 separate 
segments. In all cases, the power spectrum had a white noise at high 
frequencies, and this has been subtracted from the results shown here. 
We normalized our plots so that their integral gives the squared 
fractional rms variability, i.e. the power spectrum is in  the unit
of rms$^{2}$/Hz.

The power spectra in Figure 9, which have been averaged over all the 
observation segments, resemble the typical low-state power spectrum 
of GBHC: a flat noise at the low frequencies, rolling over to 
approximately a power law at higher frequencies (\cite{klis95}).
We find that the power spectrum varies between the energy 
bands. The IFRA (0.0025 -- 10 Hz) is $0.271\pm0.004$,  $0.349\pm0.003$,  
and $0.225\pm0.005$ for the energy bands 2 -- 5 keV, 5 -- 10 keV, 
and 10 -- 40 keV respectively. Our result is interestingly different 
from the previous {\it RXTE} observation in 1996 Aug. 3 -- 5 when the 
IFRA (0.002 -- 60 Hz) was nearly independent of energy: 26\% (2 -- 6 keV), 
27\% (6 -- 13 keV), 25\% (13 -- 28 keV) (\cite{dsmith97}). It is also 
interesting to compare our results with other  sources in their low 
states. GX 339--4, for example, had the highest variability in the 2 -- 5 keV 
band during an {\it RXTE} observation (\cite{ismith99}). This energy 
dependence of the variability may be useful in characterizing different 
X-ray sources, and its implication is explored elsewhere (\cite{ls99}).

We fit the general shape of the average PDS in Figure 9 using two zero-centered 
Lorentzians. Though a broken  power law is also commonly used in the fitting, the 
double-Lorentzian model gives better fits in our case. An additional pair of 
Lorentzians QPOs are necessary to obtain a good fit. The Lorentzian function 
is defined as:
\begin{equation}
PDS(f) = \frac{2 A f_w /\pi}{(f-f_c)^2 + f_w^2}
\end{equation}
where $f$ is the frequency, $f_c$ is the central frequency, $f_w$ is 
the half width at half maximum (HWHM), and A is the area under the 
function ($0.0 \leq f < \infty$). The fit parameters are listed in Table 8. 
Unlike the previous observations of GRS 1758--258 (\cite{dsmith97}),
the current observations detected the QPOs at 0.21 Hz and 0.42 Hz 
instead of at 0.4 Hz and 0.8 Hz.

Our power spectra are typical of low state power spectra, with two breaks at 
$\approx$ 0.1 Hz and 1 Hz. According to an inhomogeneous 
corona model (\cite{bl99}), which assumes that soft photons are injected into an 
inhomogeneous corona from a blob spiraling into the central black hole,  
the two breaks are due to different mechanisms, and they can have 
different energy dependencies. We performed  $\chi^{2}$ tests on the 
widths of the two zero-centered Lorentzians to study their dependence on energy. 
If we assume that the width of the Lorentzian is constant, a $\chi^2$ test 
on $f_{w1}$ gives a reduced $\chi^2$ of 0.67 for 2 degrees of freedom and 
a Q value of 0.51, while the test on $f_{w2}$ has a reduced $\chi^2$ 
of 3.5 for 2 degrees of freedom and a Q value of 0.021. Therefore, 
at the 95\% confidence level, the width of the lower zero-centered 
Lorentzian $f_{w1}$ has no significant difference between the three 
energy bands, but this may not be the case for the width 
of the higher Lorentzian $f_{w2}$. Further analysis using PDS ratios
shows that the general PDS shapes are energy independent for 
GRS 1758--258 (\cite{ls99b}), although this technique is less sensitive
to the detailed shape of the PDS.

In addition to studying the averaged power spectra, we also generated 
the power spectra integrated over all energies and searched for QPOs in 
each separate observation segment. A broad QPO (or ``peaked noise'') 
at $0.44 \pm 0.06$ Hz is present in all the segments. It is a very stable 
feature in the power spectrum. A pair of harmonically-spaced narrow QPOs 
at 1.4 Hz and 2.8 Hz with a total IFRA of 5\% were found in observation 
segment 7 (Figure 10). They are not  positively identified 
in the other observation segments. Only two cases of  these narrow 
significant X-ray QPOs have previously been observed in black hole candidates in 
the hard state (\cite{klis95}). 

\subsection{Phase Lags}
Another known characteristic of BHCs is that the phase lag between hard 
photons and soft photons changes very little with frequency
(\cite{miyamoto92}). This is supported by our observations (Figure 11).
Constant phase lags between the energy bands can fit the data well.
The fits give a phase lag of $(3.8 \pm 0.9) \times 10^{-2}$ radians between
the energy bands 2 -- 5 keV and 5 -- 10 keV ($\chi^2_{\nu} = 0.84 $ for 19 
degrees of freedom, Q = 0.66) and $(8.0 \pm 1.2) \times 10^{-2}$ radians 
between 2 -- 5 keV and 10 -- 40 keV ($\chi^2_{\nu} = 1.6$  for 19 degrees 
of freedom, Q = 0.042). The apparent increase of the phase
lags between the 2 -- 5 keV and 10 -- 40 keV bands may suggest a constant 
time delay over the frequencies, but the constant time delay model $\phi = 2\pi ft$ cannot 
fit the data well,  where $\phi$ is the phase lag, $f$ is the frequency, and $t$ is 
the time delay. The fit gives $\chi^2_{\nu} = 2.4$  for 19 degrees of freedom and 
Q = $5.7\times 10^{-4}$. 


This flat distribution of phase lags over a wide range of frequencies
($2\times 10^{-3}$ -- 10 Hz) was predicted by the extended atmosphere model
(\cite{kht97})
that assumes that the phase lags are generated by Compton scattering.
The atmosphere has a density
profile of $n(r) \propto \frac{1}{r}$, and the size
has to be as large as $10^4 R_{G}$, where $R_{G}$ is the Schwarzchild radius.
B\"ottcher \& Liang (1999) proposed a spiraling blob model that assumes that the seed
photons are from a blob spiraling into the central object.
It also predicted a flat phase lag distribution, but it does not need
a huge atmosphere to account for the phase lags because they
are caused by the motion of the blob from the cool outer disk
to the hot inner disk.

\subsection{Coherence \& Hardness Ratio}
 Like the previous measurements of GRS 1758--258 
(\cite{dsmith97}) and Cyg X--1 (\cite{vaughan97}; \cite{nowak99}),
the current observations found a high coherence between the hard and soft photons 
(Figure 12). The coherence between 2 -- 5 keV 
and 5 -- 10 keV is $1.02\pm0.03$, and $0.99\pm 0.04$ between 2 -- 5 keV and 
10 -- 40 keV over the frequency range of 0.01 -- 1.0 Hz.  
As \cite{hkt97} pointed out, the high coherence ($\approx 1$) 
between the photon fluxes at different energies can be produced by 
the extended atmosphere model (Hua \& Titarchuk 1995). The spiraling blob
model also predicted such a high coherence.

The hardness ratio vs intensity plots for observation segment 7 are shown in Figure 13. 
Similar results are  seen in the other segments. The hardness ratio  has little dependence
on the photon flux, although the poor statistics prevent us from drawing any significant 
conclusions. The fact that the hardness ratios show little dependence on the time 
or brightness indicates that the spectral shape we have found did not change 
markedly in 1997 August.

\section{Summary}

Our simultaneous observations of GRS 1758--258 in radio, millimeter, X-ray 
and gamma-ray add support to the idea that the source is a black hole. 
It has a very hard spectrum with a photon power law index of 1.40, 
but there is a cutoff energy around 200 keV. For Compton scattering models, 
the electron temperature is about 45 keV without considering relativistic 
effects, and 52 keV with relativistic corrections. At a distance of 8.7 kpc, 
the hard X-ray ($20 - 200$ keV) luminosity is $3.09 \times 10^{37}$ ergs/s,
which is higher than the upper limit for a neutron star binary (NSB). 
The power spectra resemble those of the BHCs, flat at low frequencies 
and breaking into a power law at higher frequencies. The phase lags 
between the hard and soft photons have a canonical frequency dependence 
for BHCs (\cite{miyamoto92}). A high coherence of about 1.0 is also found 
among the different energy bands. The flat distribution of phase lags and 
the high coherence were predicted by the extended atmosphere model and 
the spiral blob model. The radio flux was stable during the observations, 
and had a flat energy spectrum.  The millimeter observations did not 
reveal any conclusive signatures of an interaction between the jet from 
GRS 1758--258 and the molecular cloud that lies  in the direction of 
GRS 1758--258.

The spectrum can be fit with ST, (HT + BB), and (PLE + BB) models. 
The hydrogen column density deduced from the ST fitting is 
$2.0 \times 10^{22}$ cm$^{-2}$, $1.4 \times 10^{22} $ cm$^{-2}$ from
(HT + BB), and $0.93 \times 10^{22}$ cm$^{-2}$ from (PLE + BB).
It is not surprising that the value of the hydrogen column density
is model dependent  because we have valid data only down to 2.5 keV.
Though the presence of a soft component is also model dependent -- the (HT + BB)
and  (PLE + BB) models need a soft component while 
the ST model does not -- we can certainly conclude that the flux of
the soft component is low. The low flux of the soft 
component implies that the soft photon injection regions are either
small or mostly covered by the hot corona. 
One interesting result from the HT model fitting is that
the seed photons and the soft component may have different origins.
The soft component is usually believed to  originate from the
cool optically thick disk (e.g. \cite{liang98}). The seed photons may 
have two origins. One is the cool optically thick  disk, and the 
other is the internal emission processes in the corona such as 
synchrotron emission (e.g. \cite{esin96}). Therefore, the seed photons 
are very likely from the internal sources.

We  find that GRS 1758--258 is more variable in the 5 -- 10 keV energy band than
in the other two energy bands (2 -- 5 keV and 10 -- 40 keV). This property 
is interestingly different from other sources, such as GX 339--4. Systematic 
studies show that the property is indeed significantly different for the 
four hard X-ray sources: Cygnus X--1, GX 339--4, GRS 1758--258 and 1E 1740.7--2942
(\cite{ls99}). However, further observations of the sources are 
necessary to correctly understand the energy dependence of the variability 
when the sources are in different flux and spectral states.

\acknowledgements{
This work was supported by NASA grant  NAG 5-3824 at Rice University.
We thank Jean Swank, Evan Smith and the RXTE team for optimizing our 
observation plan. We thank Bob Kinzer for helping us with the 
estimation of the Galactic diffuse emission during the OSSE observations.
JM acknowledges the hospitality of the Service d'Astrophysique
at the CEA/Saclay (France), where most of the radio work was carried out,
and gives thanks to postdoctoral fellowships of the Spanish MEC and the
French CIES. JM is also partially supported by DGICYT (PB97-0903) and 
Junta de Andaluc\'{\i}a (Spain). We thank Bertrand Lefloch for 
assistance with the IRAM observations. Finally, we also thank the referee
for the valuable comments and suggestions.

The National Radio Astronomy Observatory is a facility of the National
Science Foundation operated under cooperative agreement by Associated
Universities, Inc. IRAM is funded by the Centre National de la Recherche 
Scientifique in France, the Max-Planck-Gesellschaft in Germany and the 
Instituto Geografico Nacional in Spain. The Swedish-ESO Sub-millimeter 
Telescope is operated by the Swedish National Facility for Radio Astronomy, 
Onsala Space Observatory at Chalmers University of Technology, and by ESO.

}

\clearpage

\begin{figure}[p]
\epsfysize=12cm
\epsffile{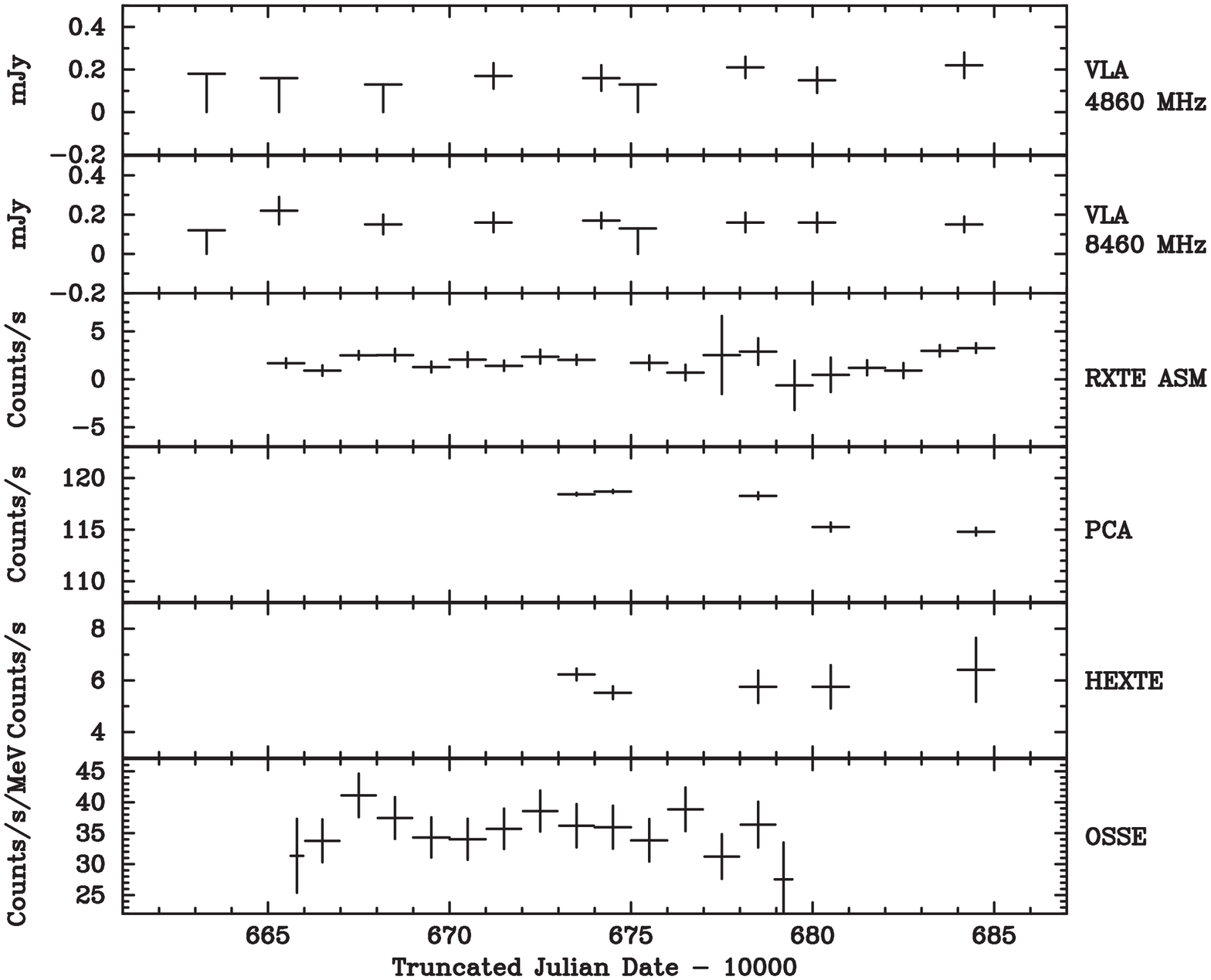}
\caption[]{The daily light curves of the VLA (4860 MHz and 8460 MHz), 
{\it RXTE} ASM (2 -- 12 keV), PCA (2.5 -- 40 keV), HEXTE (15 -- 125 keV), 
and OSSE (50 -- 100 keV) observations. We fit a constant to the light 
curves, and the fit results are given in Table 4. The one-sided crosses 
in the VLA data represent the flux density $4\sigma$ upper limits.} 
\end{figure}
\eject

\begin{figure}[p]
\begin{center}
\epsfysize=16cm
\hspace{0cm}\epsffile{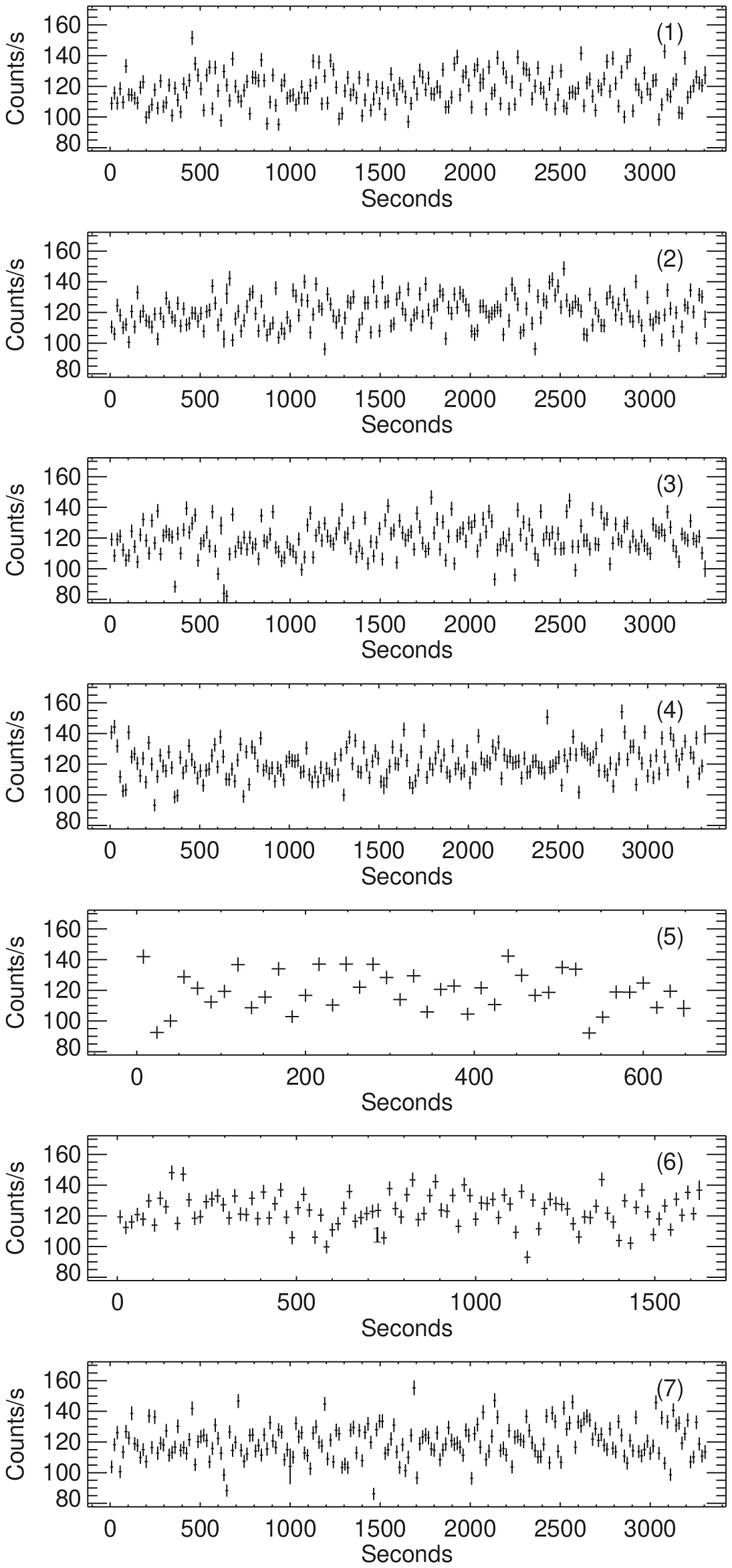}
\epsfysize=16cm
\hspace{0cm}\epsffile{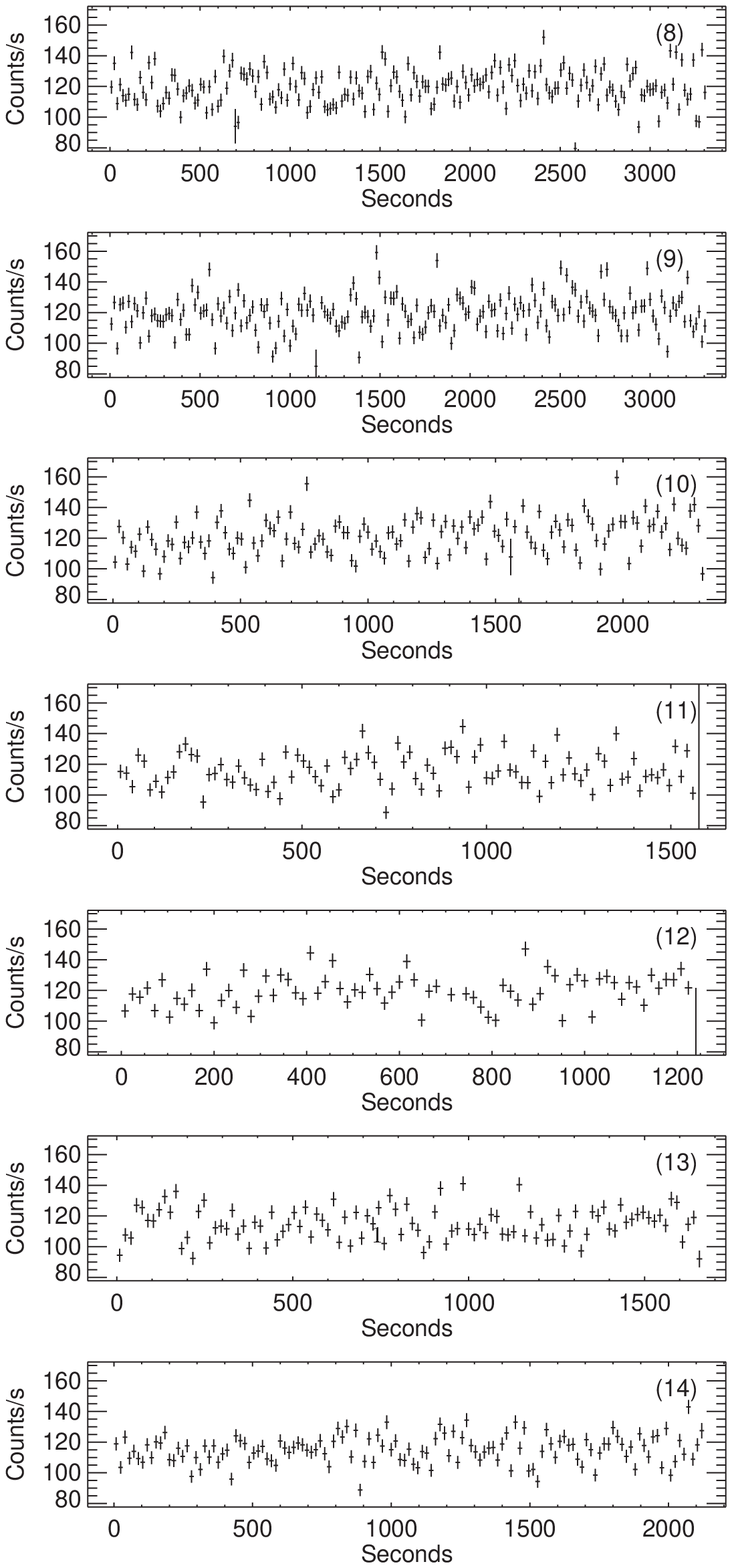}
\end{center}
\caption[]{The 16-second time resolution light curves of the 14
valid PCA observation segments. The x-axis is the time since 
the start of each observation segment, which is listed in Table 3.}

\end{figure}
\eject

\begin{figure}[p]
\begin{center}
\epsfysize=16cm
\hspace{0cm}\epsffile{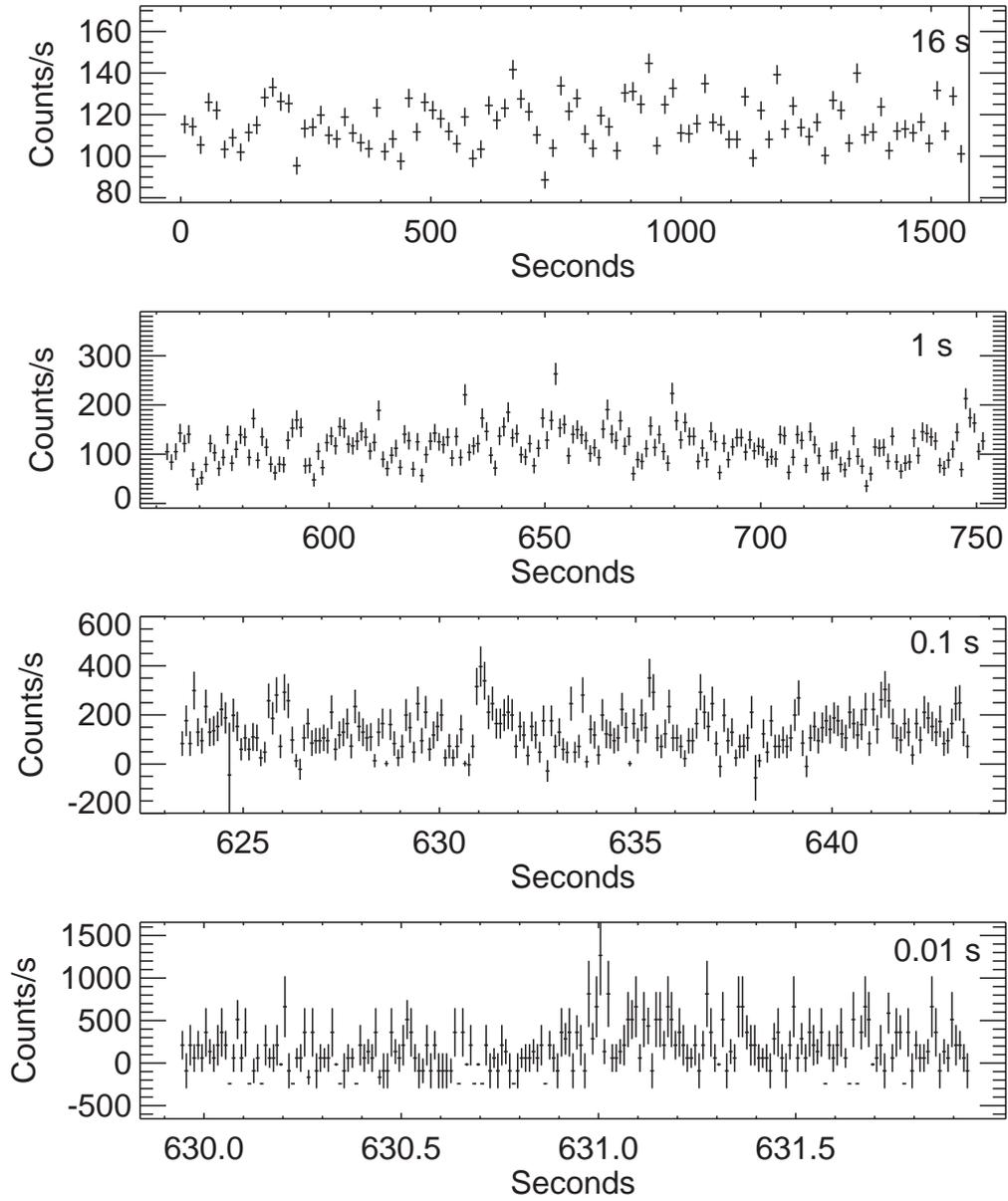}
\end{center}
\caption[]{The 16-, 1-, 0.1-, and 0.01-second time resolution 
light curves of the observation segment 11 taken with the PCA. 
The peak around 630 seconds is shown in the four time scales.}
\end{figure}
\eject

\begin{figure}[p]
\epsfysize=12cm
\epsffile{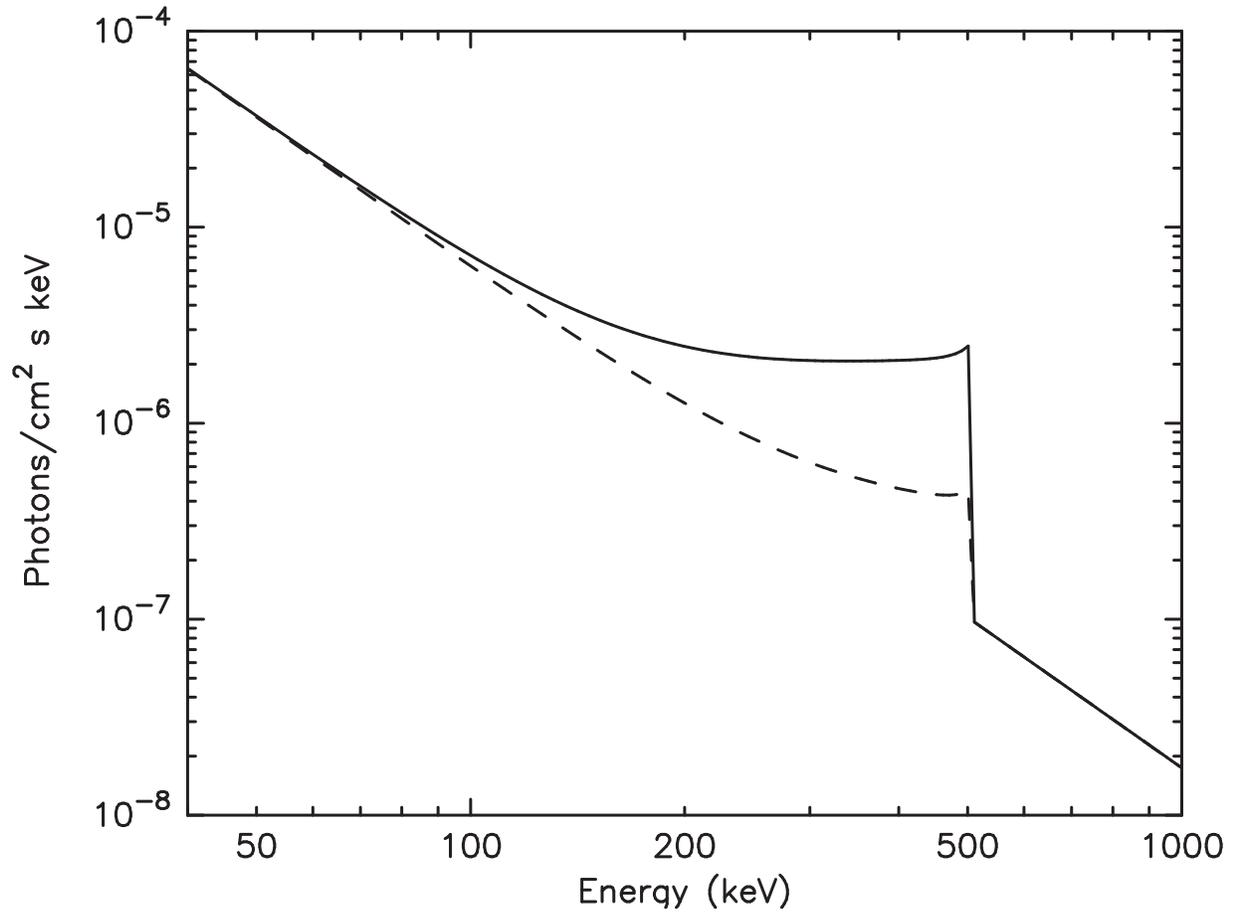}
\caption[]{The limiting versions for the background correction 
required for the OSSE data to account for the spatial variation of the GDE.
The solid line is the upper limit to the correction, and the dashed line 
is the lower limit.}
\end{figure}
\eject

\begin{figure}[p]
\epsfysize=12cm
\epsffile{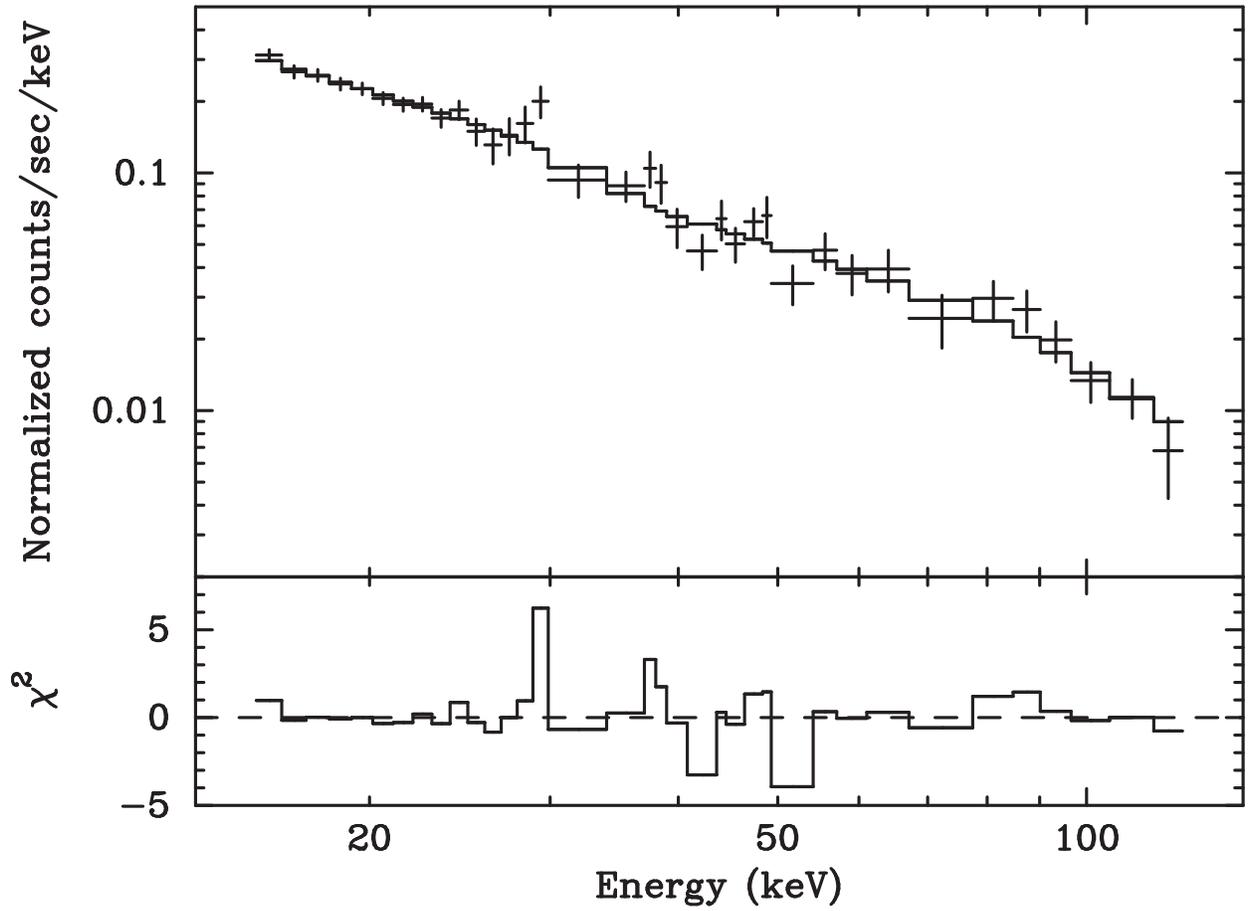}
\caption[]{A power law fit to the HEXTE cluster A data. The power 
law photon index is $1.55\pm0.4$, and the normalization at 1 keV is 
$ (9.8\pm0.13)\times 10^{-2} $ photon ${\rm cm^{-2} s^{-1} keV^{-1}}$.
The reduced $\chi^{2}$ is 1.04 for 104 degrees of freedom.} 
\end{figure}
\eject

\begin{figure}[p]
\epsfysize=12cm
\epsffile{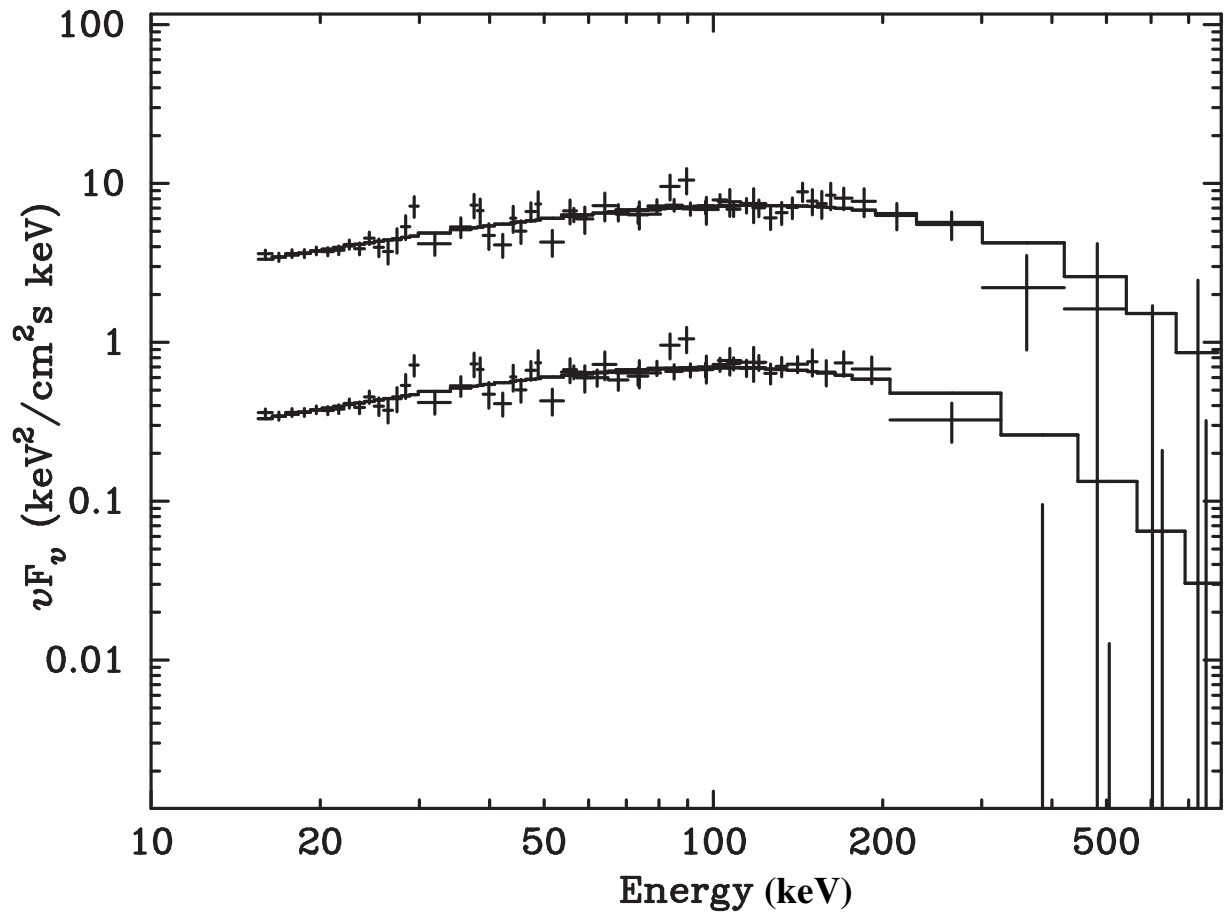}
\caption[]{The joint $\nu f_{\nu}$ spectra of the OSSE and HEXTE 
data with the two extreme OSSE background corrections. The fit 
model is a PLE. The OSSE data have been scaled up by a factor of 
1.95 to normalize to the HEXTE data. The upper curve, which has
been scaled by a factor of 10 for illustrative purposes, is the 
minimum-correction case, and has $\alpha = 1.32 \pm 0.05$ and  
$kT=179\pm 25$ keV ($\chi_{\nu}^{2} = 0.97$ for 549 degrees of 
freedom). The lower curve is the maximum-correction case, and
has $\alpha=1.26 \pm 0.06$ and $kT=140\pm 20$ keV 
($\chi_{\nu}^{2} = 0.98$ for 549 degrees of freedom).}
\end{figure}
\eject

\begin{figure}[p]
\epsfysize=12cm
\epsffile{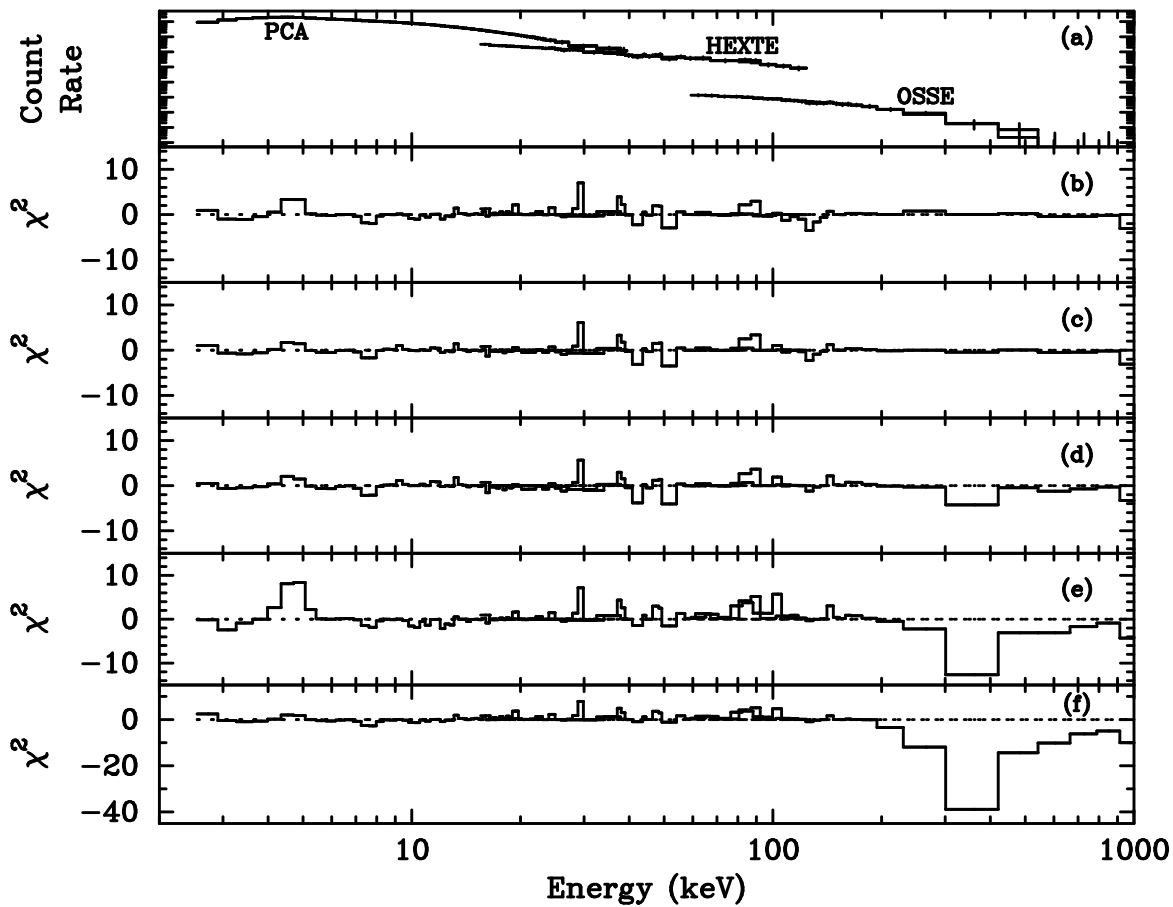}
\caption[]{Joint fits to the PCA, HEXTE and OSSE data with the ST, 
(HT + BB), (PLE+BB), PLE, and power-law models. Panel (a) is 
the normalized count spectrum using the ST model. 
The best fit model parameters are listed in Table 5. Panel (b) 
is the residual using the ST model.  Panel (c) is the residual 
using the (HT + BB) model. The best fit model parameters are 
listed in Table 6. Panel (d) is the residual using the (PLE + BB) 
model. The best fit model parameters are listed in Table 7.  Panel 
(e) is the residual from a PLE model without a soft component, 
which has $\alpha$ = $1.58\pm 0.01$, kT = $538\pm 76$ keV 
($\chi_\nu^2$ = 1.0 for 620 degrees of freedom and Q = 0.47). 
Panel (f) is the residual using a power-law model,  which has 
photon index $\alpha$ = $1.45\pm 0.01$ ($\chi_\nu^2$ = 1.19 for 
621 degrees of freedom and Q = $7.3 \times 10^{-4}$). We include 
the ISM absorption in all five cases. The PCA data has been scaled 
up by a factor of 1.33 to normalize to the HEXTE data, and the 
minimum background correction has been applied to the OSSE data.}
\end{figure}
\eject

\begin{figure}[p]
\epsfysize=12cm
\epsffile{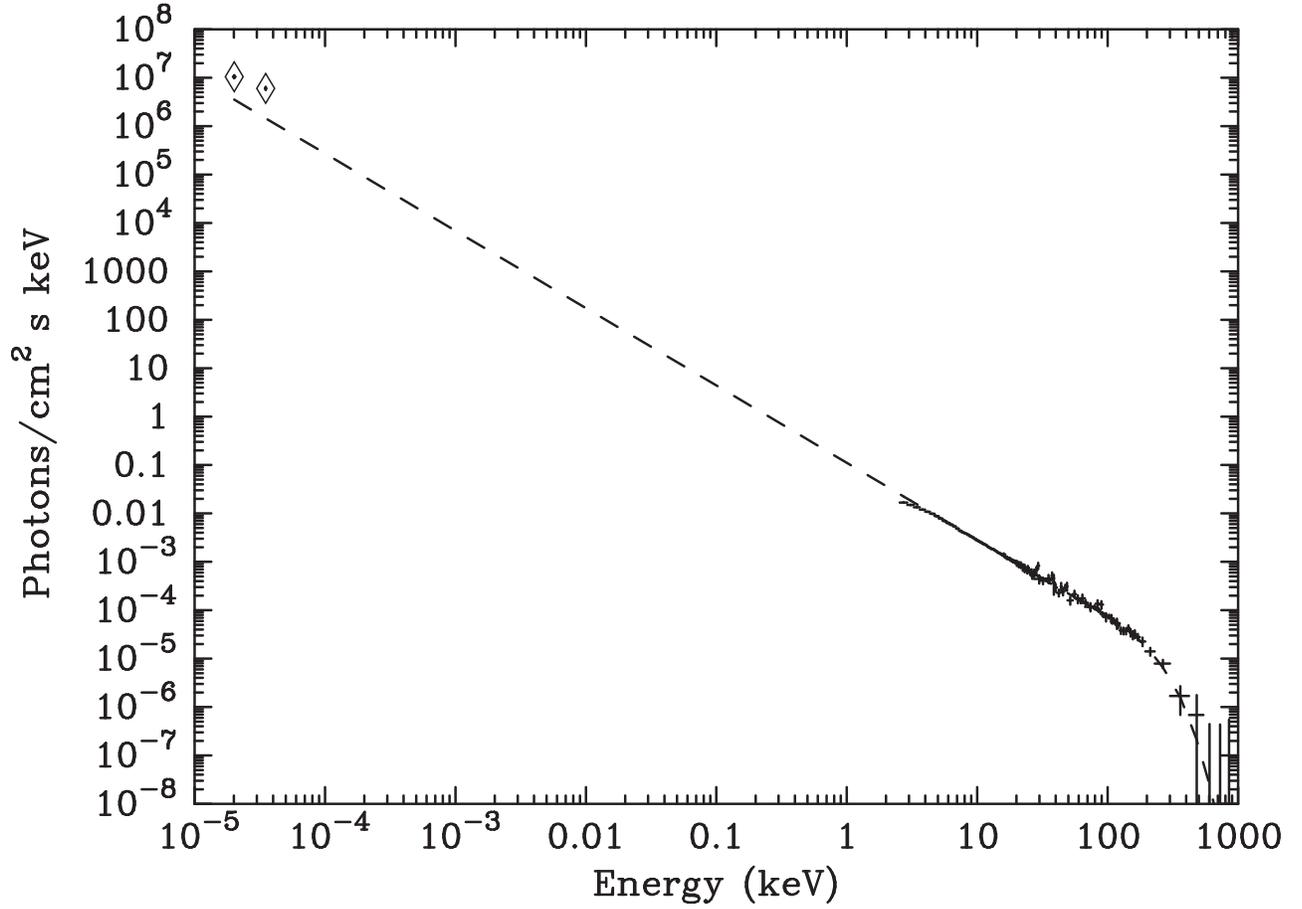}
\caption[]{The joint photon spectrum of the VLA, PCA, HEXTE and 
OSSE data. The two diamond points represent the two VLA data points. 
The dashed line is the ST model (Table 5) extrapolated to the radio 
energy without the ISM absorption. This is the same case as that 
in Figure 7 (a) \& (b). }
\end{figure}
\eject

\begin{figure}[p]
\begin{center}
\epsfysize=16cm
\hspace{0cm}\epsffile{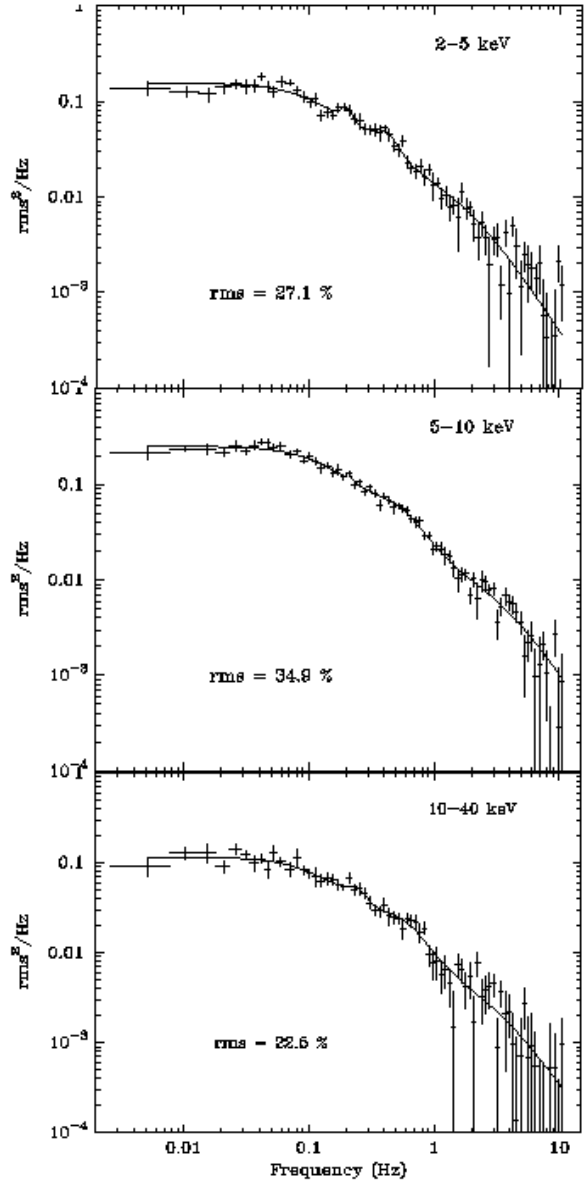}
\end{center}
\caption[]{The power spectra for the three energy bands averaged over 
all observation segments. The white noise has been subtracted.
The model includes two zero-centered
Lorentzians and a pair of harmonic QPO Lorentzians. The best fit 
parameters are listed in Table 8. The rms is calculated by summing up 
the data over the frequency range of 0.0025 -- 10 Hz.}
\end{figure}
\eject

\begin{figure}[p]
\epsfysize=12cm
\epsffile{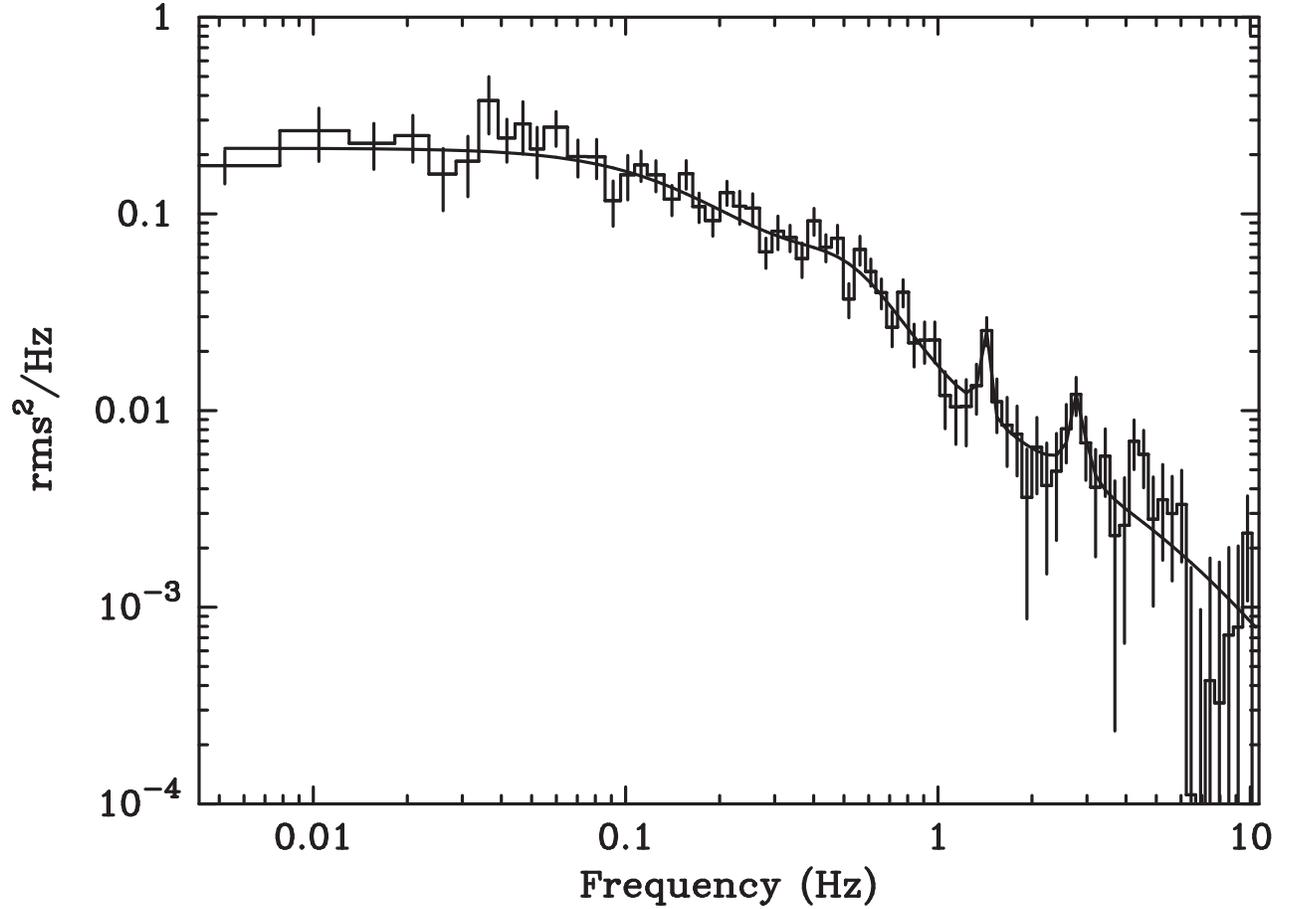}
\caption[]{The power spectrum of observation segment 7 integrated 
over all energy channels.  The white noise has been subtracted. 
The model includes two zero-centered Lorentzians, a single broad 
QPO Lorentzian at $0.44\pm 0.08$ Hz, and a pair of narrow harmonic 
QPO Lorentzians at $1.40\pm 0.04$ Hz with a width $ < 0.078$ 
Hz (2-sigma upper limit) and $2.80 \pm 0.08$ Hz with a width 
$0.12\pm 0.11$ Hz. Including the pair of narrow QPOs in the model 
increases the Q value from 0.038 to 0.28.}
\end{figure}
\eject

\begin{figure}[p]
\begin{center}
\epsfysize=12cm
\hspace{0cm}\epsffile{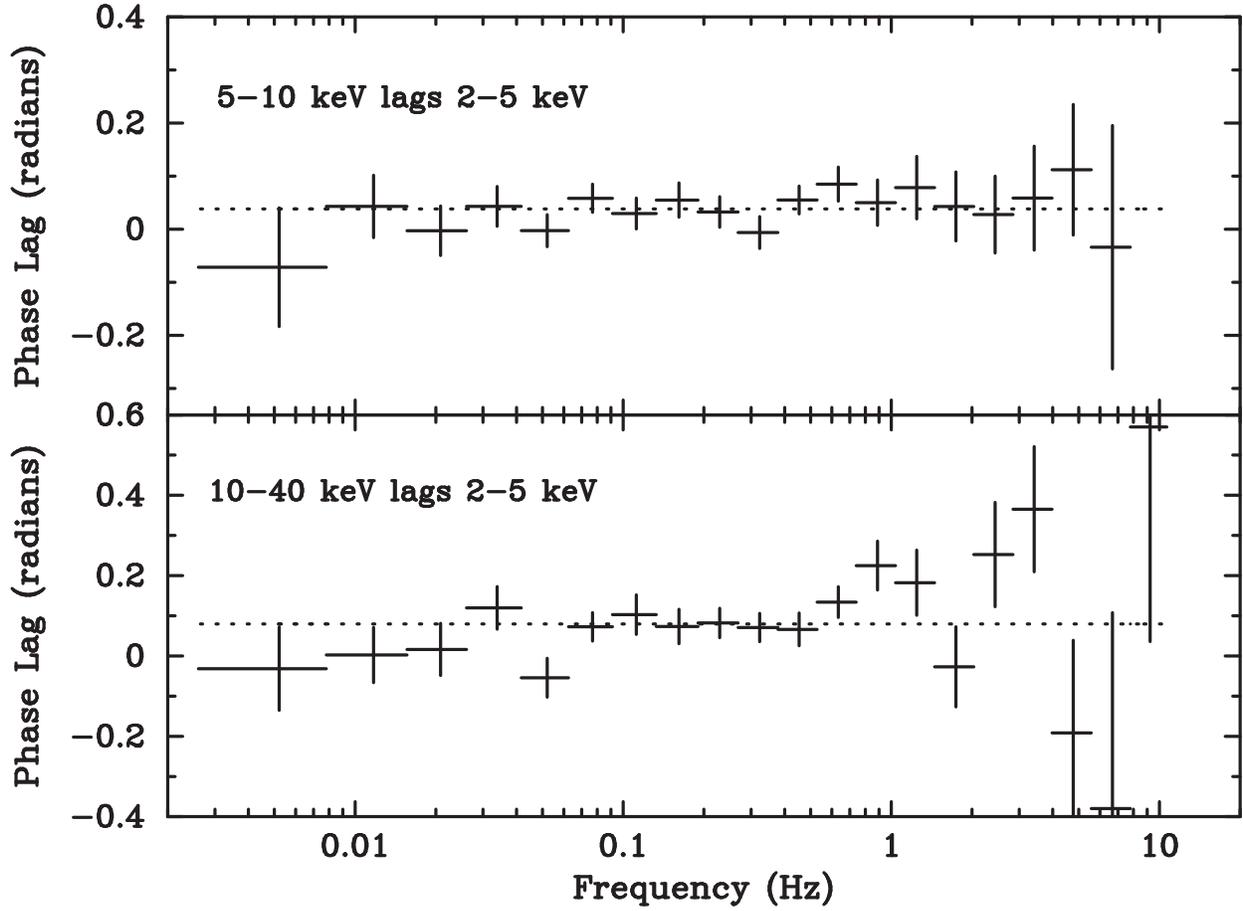}
\end{center}
\caption[]{The phase lags between the three energy bands. A positive
lag means a time delay for the hard photons. The phase lags have 
been averaged over all observation segments. The data gaps in each 
segment have been filled with Poisson noise. The dot lines are the 
weighted average phase lags.}
\end{figure}
\eject

\begin{figure}[p]
\begin{center}
\epsfysize=16cm
\hspace{0cm}\epsffile{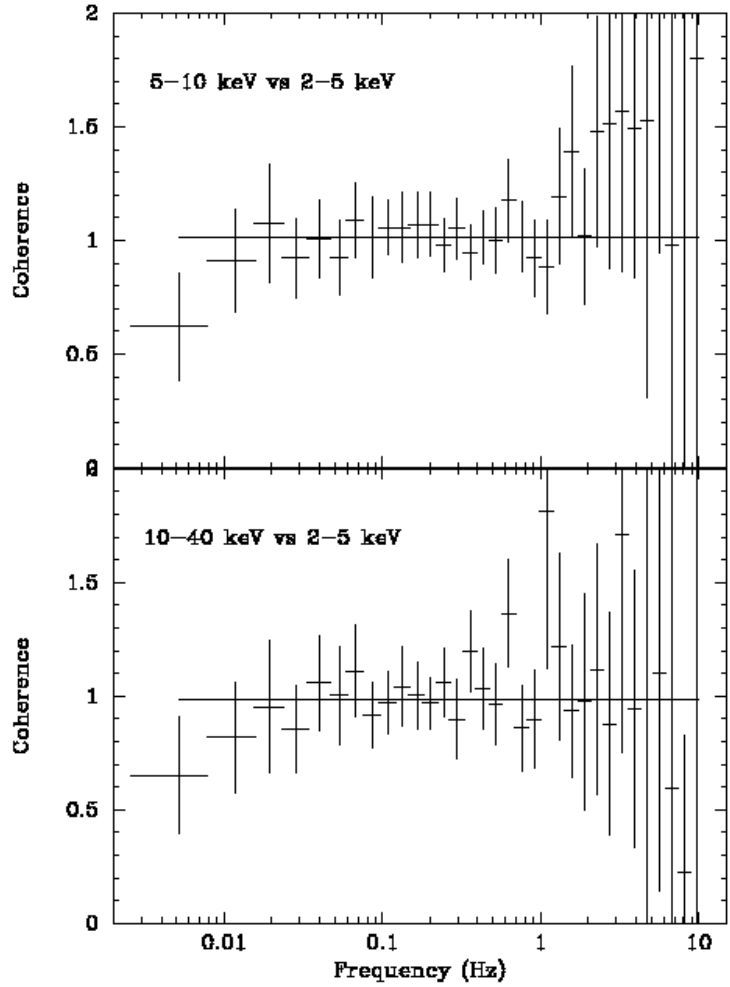}
\end{center}
\caption[]{The coherence between the three energy bands. The coherence
is calculated by averaging over all observation segments with the white 
noise subtracted.}
\end{figure}
\eject

\begin{figure}[p]
\begin{center}
\epsfysize=16cm
\hspace{0cm}\epsffile{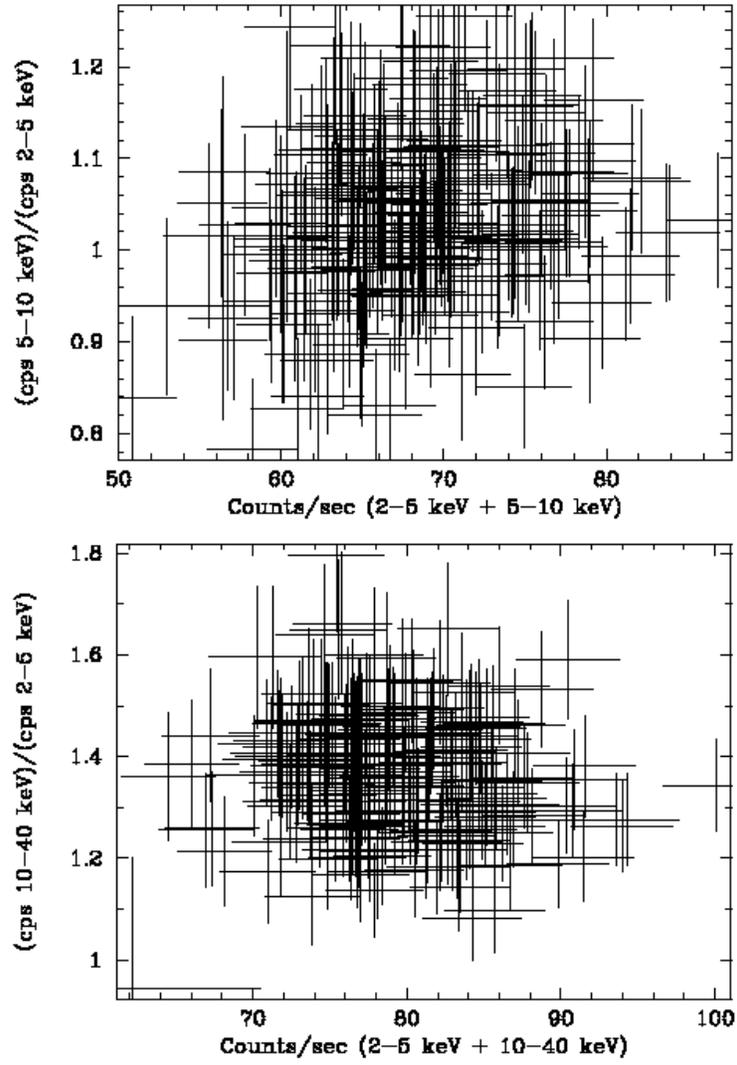}
\end{center}
\caption[]{Hardness ratios versus summed intensities for observation segment 7.
The top panel uses 2 -- 5 keV and 5 -- 10 keV. The bottom panel uses
2 -- 5 keV and 10 -- 40 keV.}
\end{figure}
\eject

\begin{deluxetable}{cll}
\tablewidth{0pt}
\tablecaption{Flux densities for VLA-C in 1997 August.}
\tablehead{
\colhead{UT Date 1997}
& \colhead{4860 MHz flux}
& \colhead{8460 MHz flux} \nl
\colhead{August (central)}
& \colhead{density (mJy)}
& \colhead{density (mJy)}}
\startdata
03.313  &  $< 0.18$          &  $< 0.12$        \nl
05.308  &  $< 0.16$          &  $0.22 \pm 0.07$ \nl
08.176  &  $< 0.13$          &  $0.15 \pm 0.05$ \nl
11.211  &  $0.17 \pm 0.06$   &  $0.16 \pm 0.05$ \nl
14.182  &  $0.16 \pm 0.06$   &  $0.17 \pm 0.04$ \nl
15.187  &  $< 0.13$          &  $< 0.13$        \nl
18.149  &  $0.21 \pm 0.05$   &  $0.16 \pm 0.05$ \nl
20.123  &  $0.15 \pm 0.06$   &  $0.16 \pm 0.05$ \nl
24.174  &  $0.22 \pm 0.06$   &  $0.15 \pm 0.04$ \nl
\enddata
\end{deluxetable}
\eject

\begin{deluxetable}{lllcclr}
\tablewidth{0pt}
\tablecaption{1995 SEST Results.}
\tablehead{
\colhead{Location}
& \colhead{Transition}
& \colhead{Velocity}
& \colhead{Antenna temp}
& \colhead{Flux density}
& \colhead{Line width}
& \colhead{Peak area}\nl
\colhead{}
& \colhead{}
& \colhead{(${\rm km~s}^{-1}$)}
& \colhead{(K)}
& \colhead{(mJy)}
& \colhead{(${\rm km~s}^{-1}$)}
& \colhead{(${\rm K~km~s}^{-1}$)}}
\startdata
Core      & \mol{12}{CO}{1-0} & $73.9 \pm 0.8$ & $1.4 \pm 0.2$  & 
$540 \pm 77$ & $10.7 \pm 1.9$ & $16.0 \pm 2.4$ \nl
          & \mol{12}{CO}{2-1} & $75.4 \pm 0.5$ & $0.7 \pm 0.25$ & 
$362 \pm 135$ & $11.3 \pm 1.1$ & $8.1 \pm 0.7$ \nl
North jet & \mol{12}{CO}{1-0} & $74.0 \pm 0.9$ & $1.2 \pm 0.2$  & 
$447 \pm 77$ & $10.1 \pm 2.0$ & $12.6 \pm 1.5$ \nl
South jet & \mol{12}{CO}{1-0} & $73.9 \pm 0.7$ & $1.6 \pm 0.2$  & 
$606 \pm 77$ & $15.0 \pm 1.8$ & $25.0 \pm 2.8$
\enddata
\end{deluxetable}

\clearpage

\newpage

\begin{deluxetable}{lllr}
\tablewidth{0pt}
\tablecaption{The start and the stop times of the 14 {\it RXTE} observation segments in Figure 2.}
\tablehead{\colhead{Segment \#} & \colhead{Start time} & \colhead{Stop time} & \colhead{TJD} 
\nl  \colhead{}  & \colhead{UT Date August 1997 } &   \colhead{UT Date August 1997} & \colhead{} }
\startdata
1  &  13.501    &   13.540   &  10673 \\

2  &  13.568    &   13.606   &  10673 \\

3  &  13.635    &   13.673   &  10673 \\

4  &  13.701    &   13.740   &  10673 \\

5  &  13.768    &   13.776   &  10673 \\

6  &  13.788    &   13.807   &  10673 \\

7  &  14.502    &   14.540   &  10674 \\

8  &  14.568    &   14.607   &  10674 \\

9  &  14.635    &   14.673   &  10674 \\

10 &  14.702    &   14.729   &  10674 \\

11 &  18.103    &   18.121  &  10678 \\

12 &  18.170    &   18.184   &  10678 \\

13 &  20.103    &   20.123   &  10680 \\

14 &  24.172    &   24.192   &  10684 \\

\enddata
\end{deluxetable}
\eject

\begin{deluxetable}{llllr}
\tablewidth{0pt}
\tablecaption{Constant fits to the daily light curves in Figure 1.}
\tablehead{
\colhead{Data set} 
& \colhead{Constant  \tablenotemark{a} } 
&\colhead{$\chi_{\nu}^2$} 
& \colhead{DOF} 
& \colhead{Q \tablenotemark{b}} }
\startdata

VLA 4860 MHz & 0.123   & 2.28    &  8   & 0.02 \\

VLA 8460 MHz & 0.139   & 1.57   &   8  & 0.13 \\

{\it RXTE} ASM      & 1.93     & 1.59   &  16 & 0.062\\

PCA                & 118.4   & 34.8   &   4 & 4.2$\times 10^{-29}$\\

HEXTE            & 5.88    & 1.156  &   4 & 0.33\\

OSSE              & 35.6   & 0.66    & 14  & 0.83\\

\enddata
\tablenotetext{a}{The constants have the same units as in Figure 1.}
\tablenotetext{b}{Q is the null hypothesis probability.}
\end{deluxetable}
\eject

\clearpage

\begin{deluxetable}{lllr}
\tablewidth{0pt}
\tablecaption{The best fit ST model to the spectrum in Figure 7(a) \& (b).}
\tablehead{
\colhead{} & \colhead{Minimum} &
\colhead{Maximum} & \colhead{ Average}}
\startdata

$n_{H}(10^{22}$ cm$^{-2}$) \tablenotemark{a} 
   & $2.0\pm 0.1$     & $2.0\pm 0.1$     & $2.0$  \\

$T_{e}$(keV) \tablenotemark{b} 
   & $48\pm 3$        & $42\pm 2$        & $45 $  \\

$\tau_{T}$ \tablenotemark{c} 
   & $3.1\pm 0.1$     & $3.4\pm 0.1$     & $3.3$  \\

$K $ \tablenotemark{d} 
   & $0.129\pm 0.004$ & $0.129\pm 0.002$ & $0.129$\\

$\chi^{2}$ (DOF)                            
   & 0.91 (620)       & 0.90 (620)       &        \\

Q & 0.94              & 0.95             &        \\

\enddata
\tablenotetext{a}{$n_{H}$ is the hydrogen column density.}
\tablenotetext{b}{$T_{e}$ is the electron temperature.}
\tablenotetext{c}{$\tau_{T}$ is the Thomson depth.}
\tablenotetext{d}{$K$ is the normalization factor of the ST model 
  defined in XSPEC}
\end{deluxetable}
\eject

\clearpage

\begin{deluxetable}{lllr}
\tablewidth{0pt}
\tablecaption{The best fit HT model to the spectrum.}
\tablehead{
\colhead{} & \colhead{Minimum} &
\colhead{Maximum} & \colhead{ Average}}
\startdata

$n_{H}(10^{22}$ cm$^{-2}$)        & $1.37 \pm 0.15$  & $1.44\pm 0.16$   & $1.4$ \\

$T_{e}$(keV)                      & $57\pm 6$        & $47\pm 4$        & $52 $     \\

$\tau_{T}$                        & $3.2\pm 0.2$     & $3.6\pm 0.2$     & $3.4$     \\

$K (10^{-2})$ \tablenotemark{a}   & $2.0\pm 0.2$     & $2.6\pm 0.2$     & $2.3$     \\

$T_{BB}$(keV)\tablenotemark{b}    & $1.19\pm 0.07$   & $1.17\pm 0.08$   & $1.18$\\

$N_{BB}({\rm \frac{L_{36}}{D_{10}^2}} )$ \tablenotemark{c}
                                  & $0.66\pm 0.12$   & $0.61\pm 0.13$   & $0.63$\\

$\chi^{2}$ (DOF)                  & 0.89 (618)       & 0.89 (620)     &           \\

Q                                 & 0.98             & 0.98          &           \\

\enddata
\tablenotetext{a}{$K$ is the normalization factor of the HT model defined in XSPEC.}
\tablenotetext{b}{$T_{BB}$ is the black-body temperature.}
\tablenotetext{c}{$N_{BB}$ is the flux from the black body. ${\rm L_{36}} $ 
  is the luminosity in the unit of $10^{36} $ ergs/s, and $ {\rm D_{10}} $ 
  is the distance in the unit of 10 kpc.
}
\end{deluxetable}
\eject

\newpage
\begin{deluxetable}{lllr}
\tablewidth{0pt}
\tablecaption{The best fit (PLE + BB) model to the spectrum in Figure 7 (d).}
\tablehead{
\colhead{} & \colhead{Minimum} &
\colhead{Maximum} & \colhead{ Average}}
\startdata

$n_{H}(10^{22}$ cm$^{-2}$)  & $0.95\pm 0.22$   & $0.97\pm 0.24$   & $0.96$\\

$\alpha$ \tablenotemark{a}  & $1.41\pm 0.03$   & $1.40\pm 0.03$   & $1.40$\\

$kT$(keV) \tablenotemark{a} & $226\pm 29$      & $189\pm 22$      & $207$\\

$N_{0} $ \tablenotemark{a}  & $0.073\pm 0.006$ & $0.071\pm 0.006$ & $0.072$\\

$T_{BB}$(keV)\tablenotemark{b} & $1.27\pm 0.06$ & $1.27\pm 0.06$  & $1.27$\\

$N_{BB}({\rm \frac{L_{36}}{D_{10}^2}} )$ \tablenotemark{b}
                               & $0.93\pm 0.23$ & $0.93\pm 0.25$  & $0.93$\\

$\chi^{2}$ (DOF)               & 0.91 (618)     & 0.92 (618)       & \\

Q & 0.95 & 0.92&\\

\enddata
\tablenotetext{a}{PLE model defined in Equation (2).}
\tablenotetext{b}{See Table 6 for explanation.}
\end{deluxetable}
\eject

\newpage
\begin{deluxetable}{lllr}
\tablewidth{0pt}
\tablecaption{The best fit model parameters to the power spectra in Figure 9.
} 
\tablehead{
\colhead{} & \colhead{2 -- 5 keV} &
\colhead{5 -- 10 keV} & \colhead{10 -- 40 keV}}
\startdata

$A_1 ({\rm rms^{2}})$ \tablenotemark{a}   & $(2.8\pm 0.3)\times 10^{-2}$ &
          $(5.4\pm 0.5)\times 10^{-2}$    & $(2.2\pm 0.3)\times 10^{-2}$ \\

$f_{w1} ({\rm Hz})$ \tablenotemark{a}     & $(1.29\pm 0.16)\times 10^{-1}$ & 
         $(1.52\pm 0.15)\times 10^{-1}$   & $(1.3\pm 0.2)\times 10^{-1}$   \\

$A_{2} ({\rm rms^{2}})$ \tablenotemark{a} & $(3.7\pm 0.3)\times 10^{-2}$   & 
         $(4.9\pm 0.4)\times 10^{-2}$     & $(1.8\pm 0.4)\times 10^{-2}$   \\

$f_{w2} ({\rm Hz})$ \tablenotemark{a}     & $1.6\pm 0.2$                  & 
         $ 3.3 \pm 0.6$                   &  $ 2.9\pm 1.6$                \\

$A_{3} ({\rm rms^{2}})$ \tablenotemark{b} & $(8.9\pm 5.3)\times 10^{-4}$  &
          *** \tablenotemark{c}           & $(1.3\pm 0.9)\times 10^{-3}$ \\

$f_{w3} ({\rm Hz})$ \tablenotemark{b}     & $(1.9\pm 1.5)\times 10^{-2}$ & 
          *** \tablenotemark{c}           & $(4.9\pm 3.0)\times 10^{-2}$ \\

$f_{c3} ({\rm Hz})$ \tablenotemark{b}     & $(2.00\pm 0.05)\times 10^{-1}$ & 
        $(2.1\pm 0.3)\times 10^{-1}$      & $(2.4\pm 0.2)\times 10^{-1}$ \\

$A_{4} ({\rm rms^{2}})$ \tablenotemark{b} & $(4.0\pm 1.4)\times 10^{-3}$ & 
        $(1.9\pm 0.5)\times 10^{-2}$      & $(8.0\pm 3.7)\times 10^{-3}$ \\

$f_{w4} ({\rm Hz})$ \tablenotemark{b}     & $(1.1\pm 0.4)\times 10^{-1}$ & 
        $(3.5\pm 0.6)\times 10^{-1}$      & $(3.5\pm 1.1)\times 10^{-1}$ \\

$\chi ^{2}$(DOF)                          & 1.43 (64) & 1.31 (64) & 0.95 (64) \\

Q                                         &  0.013  &  0.049  &  0.59  \\
\enddata
\tablenotetext{a}{The model components 1 \& 2 are the zero-center 
                  Lorentzians, i.e. $f_{c1} = 0$ and $f_{c2} = 0$.}
\tablenotetext{b}{The model components 3 \& 4 are harmonic QPO Lorentzians, 
                  i.e. $f_{c4}=2f_{c3}$}
\tablenotetext{c}{*** represents that we are not able to confine the parameter.}
\end{deluxetable}
\eject

\clearpage 


\begin{thebibliography} {}

\bibitem[Bally \& Levelthal 1991]{bl91}
     Bally, J., \& Leventhal, M. 1991, Nature, 353, 234
 
\bibitem[Barret, McClintock, \& Grindlay (1996)]{barret96}
     Barret, D., McClintock, J. E., \& Grindlay, J. E., 1996, ApJ, 473, 963


\bibitem[B\"ottcher \& Liang 1999]{bl99}
     B\"ottcher, M., \& Liang, E. P., 1999, ApJ, 511, L37

\bibitem[Chen, Gehrels, \& Leventhal 1994]{cgl94}
    Chen, W., Gehrels, N., \& Leventhal, M. 1994, ApJ, 426, 586

\bibitem[Durouchoux et al. 1999]{durouchoux99}
    Durouchoux, Ph., et al. 1999, in 32nd COSPAR Scientific Assembly, Nagoya, 
    Japan, in press 

\bibitem[Esin et al. 1996]{esin96}
    Esin, A., Narayan, R., Ostriker, E., \& Yi, I. 1996, ApJ, 465, 312.

\bibitem[Gilfanov et al. 1993]{gilfanov93}
    Gilfanov, M., et al. 1993, ApJ, 418, 844

\bibitem[Grebenev, Pavlinsky, \& Sunyaev 1997]{gps97}
    Grebenev, S. A., Pavlinsky, M. N., \& Sunyaev, R. A. 1997, in Proceedings 
    2nd INTEGRAL Workshop {\it The Transparent Universe}, ESA SP-382, 183

\bibitem[Heindl \& Smith (1998)]{hs98}
     Heindl, W. A., \& Smith D. M. 1998, ApJL, 506, L35 

\bibitem[Hjellming \& Johnston (1988)]{hj88}
    Hjellming, R. M., \& Johnston, K. J. 1988, ApJ, 328, 600

\bibitem[Hua \& Titarchuk (1995, hereafter HT, or COMPTT in XSPEC)]{ht95}
     Hua, X.-M.\& Titarchuk, L. 1995, ApJ, 449, 188

\bibitem[Hua, Kazanas, \& Titarchuk (1997)]{hkt97}
     Hua, X.-M., Kazanas, D., \& Titarchuk, L. 1997, ApJ, 482, L57


\bibitem[Jahoda  et al. 1996]{jahoda96}
     Jahoda, K., Swank, J. H., Giles, A. B., Stark, M. J., Strohmayer, 
     T., Zhang, W., \& Morgan, E. H.  1996, SPIE, 2808, 59

\bibitem[Johnson  et al. 1993]{johnson93} 
         Johnson, W. N., et al, 1993, ApJS, 86, 693

\bibitem[Kazanas, Hua, \& Titarchuk 1997]{kht97}
     Kazanas, D., Hua, X.-M., \& Titarchuk, L. 1997, ApJ, 480, 735

\bibitem[Krolik \& Kallman 1983]{kk83}
     Krolik, J. H., \& Kallman, T. R. 1983, ApJ, 267, 610

\bibitem[Kuznetsov et al. 1999]{kuznetsou99}
Kuznetsov, S. I., et al. 1999, Astronomy Letters, 25, 351

\bibitem[Lepp \& Dalgarno 1996]{ld96}
     Lepp, S., \& Dalgarno, A. 1996, A\&A, 306, L21

\bibitem[Liang 1998]{liang98}
          Liang, E. P. 1998, Physics Reports, 302, 67-142

\bibitem[Liang \& Narayan 1997]{ln97}
    Liang, E. P., \& Narayan, R. 1997, in Proceedings of the Fourth 
   Compton Symposium Part one, ed. C. D. Dermer, M. S. Strickman, 
      \& J. D. Kurfess (New York: AIP), 461

\bibitem[Lin et al. 1999a, b]{ls99}
    Lin, D., Smith, I. A., Liang, E. P., B\"ottcher, M. 1999a, BAAS, 31, 708

\bibitem[Lin et al. 1999b]{ls99b}
    Lin, D., Smith, I. A., B\"ottcher, M., Liang, E. P. 1999b, submitted to ApJ

\bibitem[Mart\'{\i} et al. 1998]{marti98}
     Mart\'{\i}, J., Mereghetti, S., Chaty, S., Mirabel, I.F., Goldoni, P., 
     Rodr\'{\i}guez, L.F., 1998, A\&A, 338, L95

\bibitem[Mereghetti, Belloni, \& Goldwurm 1994]{mereghetti94}
   Mereghetti, S., Belloni, T., \& Goldwurm, A. 1994, ApJ, 433, L21

\bibitem[Mereghetti et al. 1997]{mereghetti97}
    Mereghetti, S., Cremonesi, D. I., Haardt, F., Murakami, T., Belloni, T.,  
    \& Goldwurm, A. 1997, ApJ, 476, 829

\bibitem[Mirabel et al. 1991]{mirabel91}
    Mirabel, I. F., Morris, M., Wink, J., Paul, J., \& Cordier, B. 
    1991, A\&A, 251, L43

\bibitem[Mirabel \& Rodr\'{\i}guez 1994]{mr94}
      Mirabel, I. F., \& Rodr\'{\i}guez, L. F. 1994, in Proceedings of the
      Second Compton Symposium, eds. C. E. Fichtel, N. Gehrels, \& J. P. Norris
      (New York: AIP), 413

\bibitem[Miyamoto  et al. 1992]{miyamoto92}
    Miyamoto, S., Kitamoto, S., Iga, S., Negoro, H., \& Terada, K. 1992, 
    ApJ, 391, L21

\bibitem[Nowak et al. 1999]{nowak99}
     Nowak, M. A., Vaughan, B. A., Wilms, J., Dove, J. B., \& Begelman, M. C. 
     1999, ApJ, 510, 874

\bibitem[Oka et al. 1998]{oka98}
     Oka, T., Hasegawa, T., Hayashi, M., Handa, T., \& Sakamoto, S.
     1998, ApJ, 493, 730

\bibitem[Phillips \& Lazio 1995]{pl95}
     Phillips, J. A., \& Lazio, T. J. W. 1995, ApJ, 442, L37

\bibitem[Poutanen 1998]{poutanen98}
     Poutanen, J. 1998, in Theory of Black Hole Accretion Disks 
    (Cambridge: Cambridge U. Press)

\bibitem[Press  et al. 1996] {press96}
        Press, W. H., Teukolsky, S. A., Vetterling, W. T., \& Flannery, B. P. 
       1996, Numerical Recipes in C (Cambridge:  
        Cambridge U. press)

\bibitem[Purcell et al. 1996]{purcell96}
    Purcell, W. R.,  et al. 1996, A\&AS, 120, 389

\bibitem[Rodr\'{\i}guez, Mirabel, \& Mart\'{\i} 1992]{rmm92}
       Rodr\'{\i}guez, L. F., Mirabel, I. F., \& Mart\'{\i}, J. 1992, ApJ, 401, L15

\bibitem[Rothschild  et al. 1998]{rothschild98}
           Rothschild, R. E., et al. 1998, ApJ, 496, 538

\bibitem[Skinner 1991]{shimer91}
        Skinner, G. K. 1991, Gamma-Ray Line Astrophysics, ed. 
        P. Durouchoux \& N. Prantzos (New York: AIP), 358

\bibitem[Smith  et al. 1997]{dsmith97}
        Smith, D. M., Heindl, W. A., Swank, J., Leventhal, M., Mirabel, 
        I. F., \& Rodriguez, L. F.  1997, ApJ, 489, L51

\bibitem[Smith, Heindl, \& Swank 1999]{shs99}
       Smith, D. M., Heindl, W. A., \& Swank, J. H. 1999, IAUC 7266

\bibitem[Smith \& Liang 1999]{ismith99}
           Smith, I. A., and Liang, E. P., 1999, ApJ, 519, 771

\bibitem[Smith  et al. 1999]{ismith99a}
           Smith, I. A., {\it et al.}, 1999, ApJ, 519, 762

\bibitem[Strickman  et al. 1996]{strickman96} 
          Strickman, M., Skibo, J., Purcell, W., Barret, D., \& Motch, C. 
         1996, A\&AS, 120, 217

\bibitem[Sunyaev \& Titarchuk 1980]{st80}
      Sunyaev, R. A., \& Titarchuk, L. G. 1980, A\&A, 86, 121

\bibitem[Sunyaev  et al.  1991]{sunyaev91}
      Sunyaev, R., et al. 1991, A\&A, 247, L29

\bibitem[Tanaka \& Lewin 1995]{tanaka95}
     Tanaka, Y., \& Lewin, W. H. G. 1995, in X-Ray Binaries, ed. W. H. G.
    Lewin, J. van Paradijs, \& E. P. J. van den Heuvel 
   (Cambridge: Cambridge Univ. Press), 126

\bibitem[Titarchuk (1994)]{titarchuk94}
     Titarchuk, L. 1994, ApJ, 434, 313

\bibitem[Van der Klis 1995]{klis95}
         Van der Klis, M., 1995, in X-Ray Binaries, ed. W. H. G. Lewin, J.  
         Van Paradijs, and E. P. J. Van den Heuvel(Cambridge: Cambridge   
         U. press), 252

\bibitem[Vaughan \& Nowak 1997]{vaughan97}
    Vaughan, B. A., \& Nowak, M. A. 1997, ApJ, L43

\bibitem[Yan \& Dalgarno 1997]{yd97}
    Yan, M., \& Dalgarno, A. 1997, ApJ, 481, 296


\end{thebibliography}
\end{document}